\documentclass[11pt,a4paper]{article}
\pdfoutput=1
\usepackage{jcappub}

\usepackage[T1]{fontenc}
\usepackage[utf8]{inputenc}
\usepackage{natbib}
\usepackage{amssymb,amsmath}
\usepackage[english]{babel}
\usepackage[normalem]{ulem}

\def\be{\begin{equation}}
\def\ee{\end{equation}}

\makeatletter

\usepackage{graphicx}
\usepackage{color}
\definecolor{linkblue}{rgb}{0,0,0.8}
\definecolor{linkgreen}{rgb}{0,0.5,0}
\definecolor{darkgreen}{rgb}{0,0.4,0}
\definecolor{purple}{rgb}{0.7,0.0,0.4}
\hypersetup{linkcolor=linkblue, citecolor=purple, urlcolor=linkblue}

\usepackage[usenames,dvipsnames,svgnames]{xcolor}
\usepackage{amsmath,amssymb}

\usepackage{cancel}
\usepackage{braket}

\usepackage[colorinlistoftodos]{todonotes}
\usepackage{import}

\makeatother

\begin{document}

\title{Dissipative Axial Inflation}

\author{Alessio Notari$^{1}, $}
\author{Konrad Tywoniuk$^{2}$}

\affiliation{$^{1}$   Departament de F\'isica Fondamental i Institut de Ci\`encies del Cosmos, Universitat de Barcelona, Mart\'i i Franqu\`es 1, 08028 Barcelona, Spain.}
\affiliation{$^{2}$ Theoretical Physics Department, CERN, Geneva, Switzerland. }

\begin{abstract}
{ We analyze in detail the background cosmological evolution of a scalar field coupled to a massless abelian gauge field through an axial term $\frac{\phi}{f_\gamma} F \tilde{F}$, such as in the case of an axion. Gauge fields in this case are known to experience tachyonic growth and therefore can backreact on the background as an effective dissipation into radiation energy density $\rho_R$, which which can lead to inflation without the need of a flat potential.  We analyze the system, for momenta $k$ smaller than the cutoff $f_\gamma$, including numerically the backreaction. We consider the evolution from a given static initial condition and explicitly show that, if $f_\gamma$ is  smaller than the  field excursion $\phi_0$ by about a factor of at least ${\cal O} (20)$, there is a friction effect which turns on before that the field can fall down and which can then lead to a very long stage of inflation with a generic potential. In addition we find superimposed oscillations, which would get imprinted on any kind of perturbations, scalars and tensors. Such oscillations have a period of 4-5 efolds and an amplitude which is typically less than a few percent and decreases linearly with $f_\gamma$.  We also stress that the comoving curvature perturbation on uniform density should be sensitive to slow-roll parameters related to $\rho_R$ rather than $\dot{\phi}^2/2$, although we postpone a calculation of the power spectrum and of non-gaussianity to future work and we simply define and compute suitable slow roll parameters. Finally we stress that this scenario may be realized in the axion case, if the coupling $1/f_\gamma$ to U(1) (photons) is much larger than the coupling $1/f_G$ to non-abelian gauge  fields (gluons), since the latter sets the range of the potential and therefore the maximal allowed $\phi_0\sim f_G$. }
\end{abstract}

CERN-TH-2016-189

\maketitle

\section{Introduction}
Inflation~\cite{Starobinsky:1980te,Guth:1980zm,Linde:1981mu} is the best known candidate for providing initial conditions to our Universe, which can then evolve in the hot radiation dominated era. It explains in fact how the Universe can become nearly homogenous, isotropic and flat on very large scales. Moreover a great success of inflation is the possibility of generating an initial condition for metric perturbations which are consistent with the observed spectrum on large scales: adiabatic, gaussian and almost scale invariant. The simplest realization of inflation is arguably a single slowly rolling scalar field. This requires nonetheless a very flat potential, in such a way that the absolute value of the potential dominates over its derivatives, and so the Hubble friction dominates the evolution. It usually requires a large amount of fine-tunings to explain how such a potential can maintain its flatness during the required 60 efolds of inflation. In the present paper we pursue the idea that inflation could actually be realized in a very different way, in which the slow-roll is not due to Hubble friction, but is dynamically generated by a non-trivial dissipation mechanism into other degrees of freedom.  Such an idea has been invoked already in some scenarios, {\it e.g.} in~\cite{Berera:1995ie}. In our view a simple and compelling concrete mechanism is the one proposed in~\cite{Anber:2009ua}~\footnote{ This was also called ``natural steep'' inflation, while we prefer to avoid the word natural, because it reminds necessarily of axion potentials, while the only ingredient is an axial coupling which allows dissipation into gauge fields, and so we prefer to call it dissipative axial inflation.}, in which a scalar $\phi$ is coupled to a CP-odd combination of a $U(1)$ gauge field, with the term  $\frac{\phi}{f_\gamma} F \tilde{F}$: in this case in fact a nonzero time derivative $\dot{\phi}$ induces a tachyonic instability for the gauge field, which can then grow until it backreacts on the scalar, slowing down its evolution. In the present paper we study such a system by fully taking into account of the backreaction in a numerical way, showing that indeed slow-roll can be realized even in absence of Hubble friction and studying in detail the evolution of the system.  As we will discuss, the analysis performed in~\cite{Anber:2009ua} assumed that such a friction dominated regime can be realized and estimated it in an analytical approximation. There are some issues which  however cannot be addressed within such a treatment. For instance such an estimate assumes that an inflationary solution with almost constant expansion rate $H$ and almost constant field velocity $\dot{\phi}$ can be found, but does not address how such a solution can be reached from a given initial condition. In fact the question can be formulated as follows: starting with a field at rest at some initial value in a steep potential, can the backreaction term turn on before that the field simply falls down its potential? The initial stage of evolution can be actually studied ignoring the expansion of the Universe since the field dynamics dictated by the potential is much faster than the Hubble expansion. The treatment of~\cite{Anber:2009ua} cannot be used for this purpose, since it actually explodes exponentially in the limit $H\rightarrow 0$ and only makes sense after the backreaction has already turned on. Another reason why a numerical treatment is appropriate is to check whether the ansatz of constant $\dot{\phi}$ is actually realized and to which extent.  For these reasons we study the system carefully in the present paper and show that it can be solved explicitly, therefore providing an inflationary background and finding new relevant features. Note that a similar numerical analysis has been performed in~\cite{Cheng:2015oqa}, although with the scalar starting in a Hubble friction regime and then only approaching the dissipative regime in the end of inflation. We will comment more on this study in the rest of the paper. Importantly we also find that generically the energy density dissipated in radiation $\rho_R$ is much bigger than the field kinetic energy and so it dominates the time derivative of the total energy density $\dot{\rho}$; we stress then that the amplitude of the primordial curvature perturbation $\zeta$ should be different from the usual expression for a single scalar field used in~\cite{Anber:2009ua}, although we do not perform here a full calculation of $\zeta$. Finally we will also point out that this mechanism can work with generic potentials, but most naturally can be realized in the case of an axion or axion-like field coupled to a $U(1)$ gauge field: in fact, using the well-known example of the QCD axion we suggest that the scenario can work  just by having the coupling between axion and abelian fields (the photon) much larger than than the coupling to non-abelian fields (the gluons).

The paper is organized as follows.
In section~\ref{setup} we review the setup. In section~\ref{flat} we solve the system in flat spacetime, while in section~\ref{Inflation} we solve it in an expanding background. In section~\ref{results} we show results for several potentials. Finally we discuss some  features of the model in~\ref{discussion} and we draw our conclusions in~\ref{conclusions}.
\section{Setup} \label{setup}
We consider the action of a scalar field $\phi$ with a potential $V(\phi)$, coupled to a massless gauge field $A_\mu$ (the photon or another field), with field strength $F_{\mu \nu}=\partial_\mu A_\nu-\partial_\mu A_\nu$. Such a coupling is naturally present for an axion field and it has been considered in~\cite{Anber:2009ua} for inflation.  In principle one could also consider non-abelian fields, but this would introduce complications: ($i$) the presence of nonlinear self-interactions might change the production mechanism, ($ii$) producing friction mostly through such a coupling would not work, because it would also induce a contribution to $V(\phi)$ through instanton effects; we therefore only consider U(1) and discuss later the importance of the relative size of abelian vs. non-abelian couplings.
Our action is therefore
\begin{eqnarray}
S=-\int d^4x \sqrt{-g} \left(\frac{1}{2}\partial_\mu \phi \partial^\mu \phi+ V(\phi) +\frac{1}{4} F_{\mu \nu} F^{\mu \nu} + \frac{g(\phi)}{4} F_{\mu \nu} \tilde{F}^{\mu \nu} \right)
\, , \end{eqnarray}
where we assume for simplicity the coupling function to be
\begin{eqnarray}
g(\phi)=\frac{\phi}{f_{\gamma}}
\, , \end{eqnarray}
and $f_\gamma$ is a mass scale and $\tilde{F}_{\mu \nu}$ is the antisymmetric dual field tensor. It is  straightforward to generalize to other forms of $g(\phi)$. Note that such a coupling term has dimension five and therefore we can trust it only at energy scales smaller than $f_\gamma$. We will return on this point and on the possible origin of such a coupling from a more complete renormalizable theory at higher energy in section~\ref{discussion}. 
We first analyze in section~\ref{flat} the system in flat spacetime in order to have a clear picture of the dynamics and of the onset of the backreaction. In fact we are interested in cases in which the scalar field potential is steep, $\partial V/\partial \phi\equiv V_{,\phi}\gg 3 H \dot{\phi}$, and therefore the Hubble friction is subdominant, at least in the initial stages of the $\phi$ evolution, until the backreaction term becomes important. Then we will include the expansion of the metric in section~\ref{Inflation}.

\section{Flat spacetime analysis} \label{flat}

Let us first analyze the equations in the simplest setup, namely in flat spacetime. Varying the action with respect to $\phi$ and only considering the spatially homogeneous mode $\phi(t)$ we get the equation of motion:
\begin{eqnarray}
\ddot{\phi}+V_{,\phi}(\phi)+  g_{,\phi}(\phi) \, {\cal B} =0, \qquad  {\cal B} \equiv\langle F\tilde{F}\rangle /4 
 \label{phiflat}
\, . 
\end{eqnarray}
A gauge field of three momentum $\vec{k}$ evolves in the $\phi(t)$ background according to (see {\it e.g.}~\cite{Tkachev:1986tr} and references therein):
\begin{eqnarray}
\ddot{A}_\pm+ (k^2\mp k \dot{g}) A_\pm=0 \label{Aflat}
\, , \end{eqnarray}
where $k\equiv |\vec{k}|$ and the index $+$ or $-$ represents the gauge field helicity, defined by a helicity vector $\vec{\epsilon}_\pm$, which satisfies $\vec{k} \cdot \vec{\epsilon}_\pm=0$ and $\vec{k} \times\vec{\epsilon}_\pm=\mp i k \vec{\epsilon}_\pm$.
The gauge field backreacts on the scalar through 
\begin{eqnarray}
{\cal B} =   \frac{1}{2} \int \frac{d^3k}{(2\pi)^3} \, k \frac{d \left[ |A_+|^2 - |A_{-}|^2 \right] }{dt} 
\label{FFdualflat}
\, , \end{eqnarray}
In  absence of  interactions the classical field evolution would be simply determined by the potential. The interacting system has instead a non-trivial evolution if the coupling $1/f_{\gamma}$ is large enough: there can be very efficient dissipation of energy into gauge fields and this can slow down dramatically the scalar field evolution, being effectively a friction-like term. The physical essence is that a tachyonic instability signals that the initial background is not the correct one, and by solving for the full system we effectively find a new background. Note that the existence of such a friction is possible only because our interaction term involves a CP-odd combination $F \tilde{F}$, which does not respect Time reversal (T) and therefore it is sensitive to $\dot{\phi}$, contained in $\dot{g}$. The fact that the friction can be sustained for a long time is crucially due to the massless nature of the gauge field, so that, even if the field $\phi$ changes value this does not induce a mass, which would otherwise prevent efficient dissipation.

Let us now place our field at rest at some initial time $t=0$ with an initial value $\phi_0>0$ and for simplicity consider a constant force $V_{,\phi}\equiv -V_p <0$, so that the solution without backreaction is 
\begin{eqnarray}
\phi(t)= \phi_0 -\frac{1}{2} V_p t^2 \, ,
\end{eqnarray}
and so $\phi$ would fall down its potential to a final value (introducing additional terms to the linear potential this could be a minimum of $V$, which we can also define to be $\phi=0$ without loss of generality), in a time given by $t_F\approx (\phi_0/V_p)^{1/2}$. In this case the gauge field behaves according to
\begin{eqnarray}
 \ddot{A}_{\pm} + (k^2\pm \mu^2 k  t) A_{\pm}=0\, , \qquad \mu^2\equiv \frac{V_p}{f_\gamma} \, .
\end{eqnarray}
We can focus on the negative helicity modes, which for low $k$ are unstable. We find the following solution in terms of the Airy functions of the first and second kind:
\begin{eqnarray}
A_{-}(t)= c_1 \text{Ai}\left(\frac{k \left(\mu ^2 t-k\right)}{\left(k \mu ^2\right)^{2/3}} \right)+c_2 \text{Bi}\left(\frac{k \left(\mu ^2 t-k\right)}{\left(k \mu ^2\right)^{2/3}} \right) \label{Airy}
\, , \end{eqnarray}
where $c_1$ and $c_2$ are two constants related to the initial conditions. In the large $t$ limit the second term of this equation is rapidly growing for low $k$:
\begin{eqnarray}
A_{-}\approx c_2\frac{e^{\frac{2 \sqrt{k} \left(\mu ^2 t-k\right)^{3/2}}{3 \mu^2}}}{\sqrt{\pi } \label{Aexp}
   \sqrt[12]{k} \sqrt[6]{\mu} \sqrt[4]{t}} \label{Ak}
   \, , \end{eqnarray}
More and more modes grow as time grows. At a given time all modes with $k<\mu^2 t$  grow explosively, with a time-scale given by $\bar{t}\approx (k \mu^2)^{-1/3}$.  We should then integrate eq.~(\ref{FFdualflat}) using eq.~(\ref{Ak}) for all the tachyonic $k$'s. This can be done analytically approximating $\mu ^2 t-k\approx \mu ^2 t$ in the exponents, leading to a ${\cal B}$ which grows roughly as $e^{\frac{4}{3} (t \mu)^2}$. This means that we should compare the timescale $\mu^{-1}$ with $t_F$, in order to know whether the effect can become large before that the field falls down. This gives us the crucial condition that $f_{\gamma} \gg \phi_0$. Such an analysis is valid for a constant force, but a similar  general lesson can be extracted from the more realistic (and complicated) case of a quadratic potential $V=\frac{1}{2} m^2 \phi^2$. In this case the free field solution is periodic with some amplitude $\phi_0$, so that with suitable initial conditions we have:
 \begin{eqnarray}
\dot{\phi}(t)= m \phi_0 \cos(m t) \, ,
\end{eqnarray}
which leads to
\begin{eqnarray}
 \ddot{A}_{\pm} + \left( k^2\mp  \frac{k  m \phi_0 }{f_\gamma}  \cos(m t) \right) A_{\pm}=0\,  .
 \label{flatquadratic}
\end{eqnarray}
This equation has a well-known solution
\begin{eqnarray}
 A_{\pm}(t) &=& c_1 \text{MathieuC}[a,q,z]+ c_2 \text{MathieuS}[a,q,z] \label{Mathieu} \, , \nonumber \\ 
 a &\equiv & \frac{4k^2}{m}, \,  \, \,  q \equiv \pm \frac{2 k \phi_0}{m f_\gamma}, \,  \, \,  z \equiv \frac{m t}{2} \, ,
\end{eqnarray}
 in terms of even and odd Mathieu functions, whose behavior can be always written, according to Floquet's theorem, as $e^{i r z} f(z)$, where $f(z)$ has period $2\pi$ and $r$ is the so-called Mathieu characteristic exponent. It is possible to check that  for $k< \frac{m \phi_0}{ f_\gamma}$ such exponents have a negative imaginary part for one of the two helicities and so lead to an exponential growth of the modes. Such a growth can overcome the free field classical evolution if the $|{\rm Im}(r)|$'s are much bigger than 1. Therefore one can plot ${\rm Im}(r)$ and check that, fixing for instance units of $m=1$, they grow linearly as $\kappa \frac{\phi_0}{f_\gamma}$, where $\kappa$ is a number which depends on $k$ and which has a maximum value of roughly  $0.3$. This shows again therefore that one can get large backreaction as long as $\phi_0/f_\gamma$ is  bigger than a few times $\kappa^{-1}$, which means ${\cal O}(10-20)$.
 
As a next step we solve numerically the coupled system eqs.~(\ref{phiflat}-\ref{FFdualflat}), for a given potential $V(\phi)$. We solve eq.~(\ref{Aflat}) for a relatively large number, $N_{\rm modes}\approx {\cal O}(10-100)$, of different discrete values of $k$.
We set the following initial conditions:
\begin{eqnarray}
A(0)&=&\frac{1}{\sqrt{2 k}} \, , \qquad \dot{A}(0) = \frac{i k}{\sqrt{2 k}} \, , \nonumber \\
 \phi(0)&=&\phi_0 \, , \qquad \dot{\phi}(0)= 0 \, . \label{initial}
\end{eqnarray}
The first line corresponds to vacuum fluctuations for the gauge field.
Then, we include the backreaction term in the system of differential equations by discretizing the integral in $N_{\rm modes}$ logarithmic intervals. We check that the result is not sensitive to the number of modes, as longs as we capture the relevant range. In order to find such a range we plot the backreaction spectrum, defined through 
\begin{equation}
{\cal B} \equiv \int \frac{dk}{k} P_{\cal B}(k) \label{spectrum}
\end{equation} 
and we will check that we integrate over the relevant region. 

\begin{figure}[t]
    \centering
    \includegraphics[width=0.75 \textwidth]{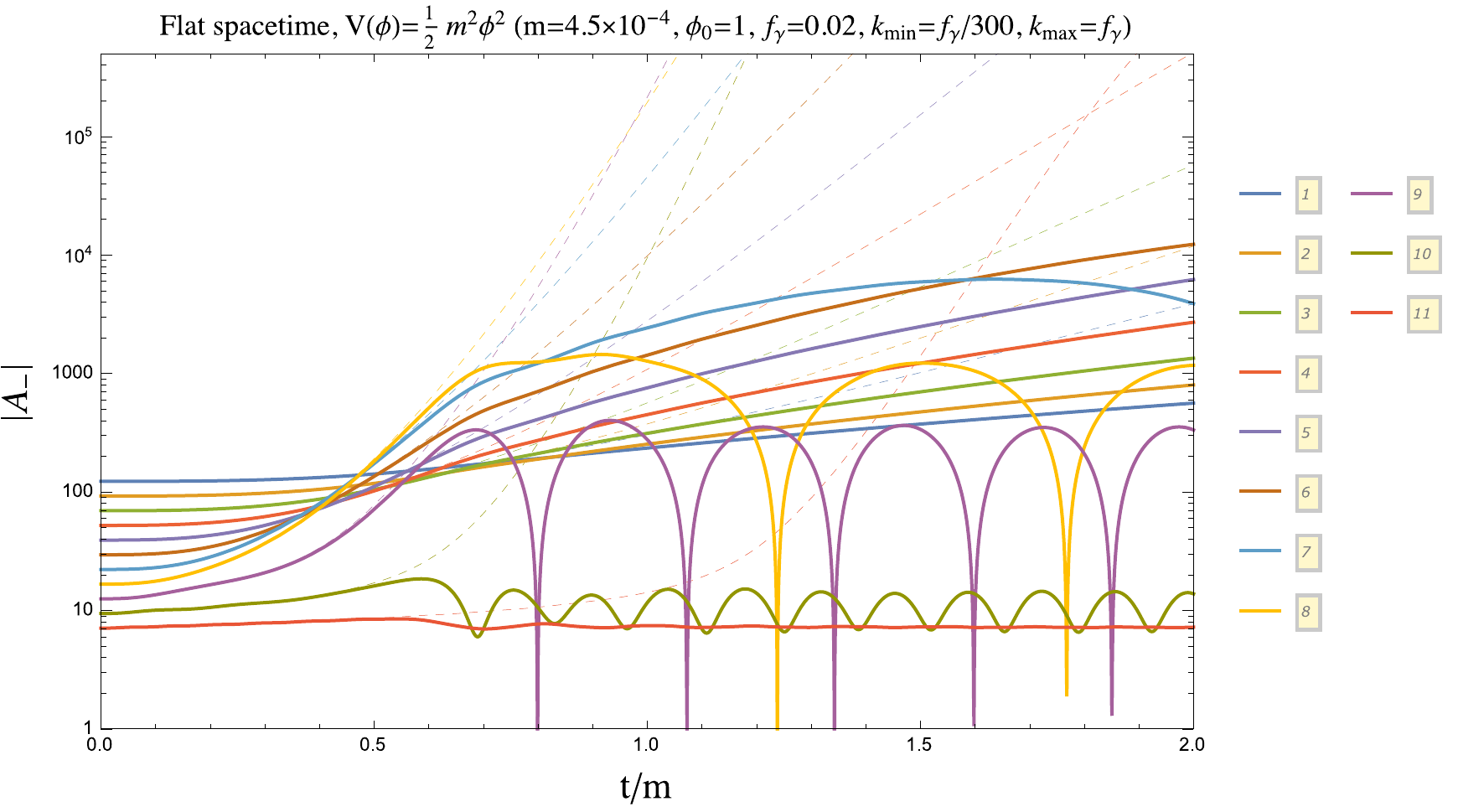}
    \caption{We show the time evolution, in a quadratic potential and in flat spacetime, of the growing gauge field $A_{-}$, for increasing values of $k$, from $k_{\rm min}=f_\gamma/300$ to $k_{\rm max}=f_\gamma$, equally spaced in logarithmic scale. In dashed lines we show the same modes, in absence of backreaction.}
    \label{Atflat}
\end{figure}

\begin{figure}[t]
    \centering
    \includegraphics[width=0.46 \textwidth]{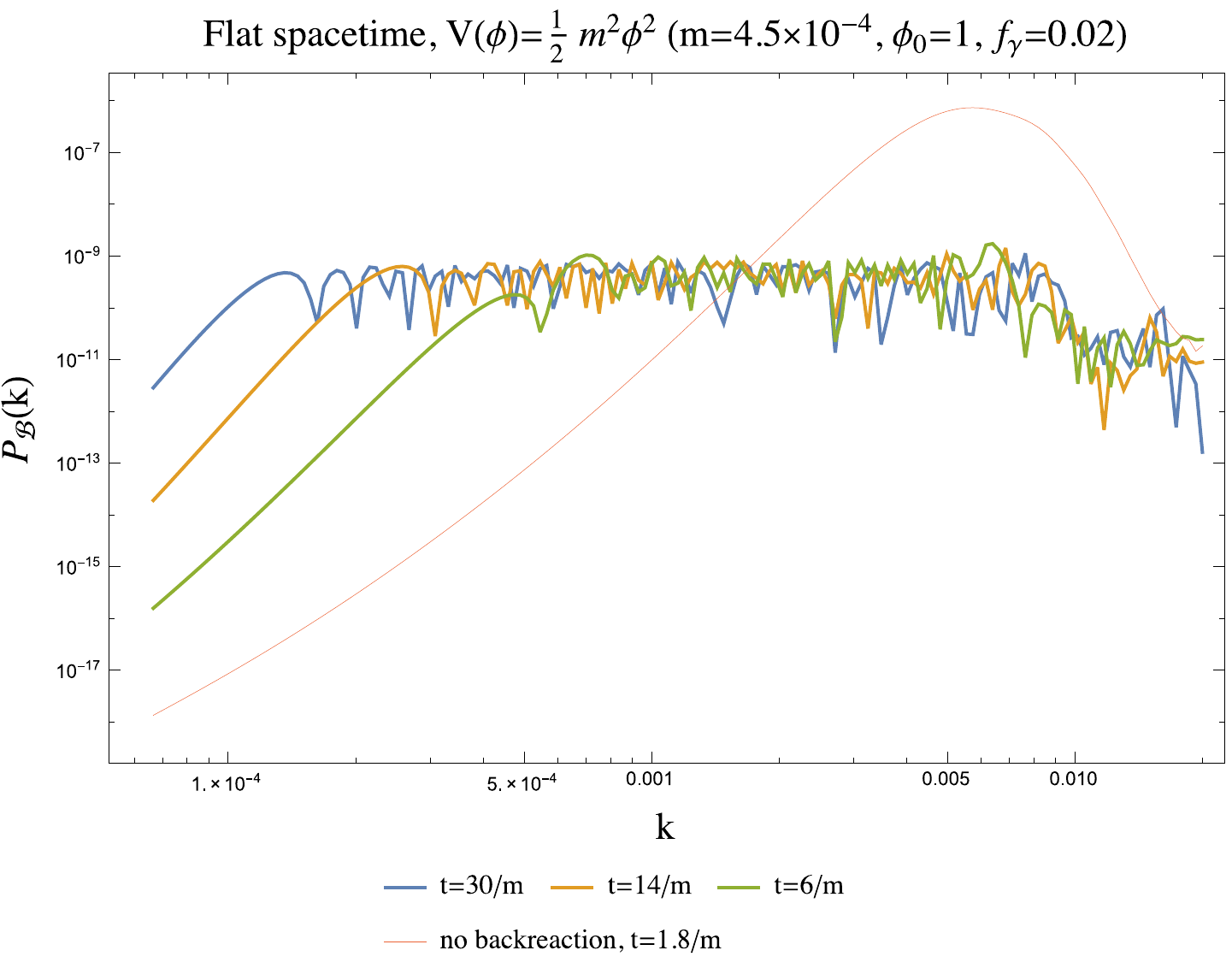}
    \includegraphics[width=0.46 \textwidth]{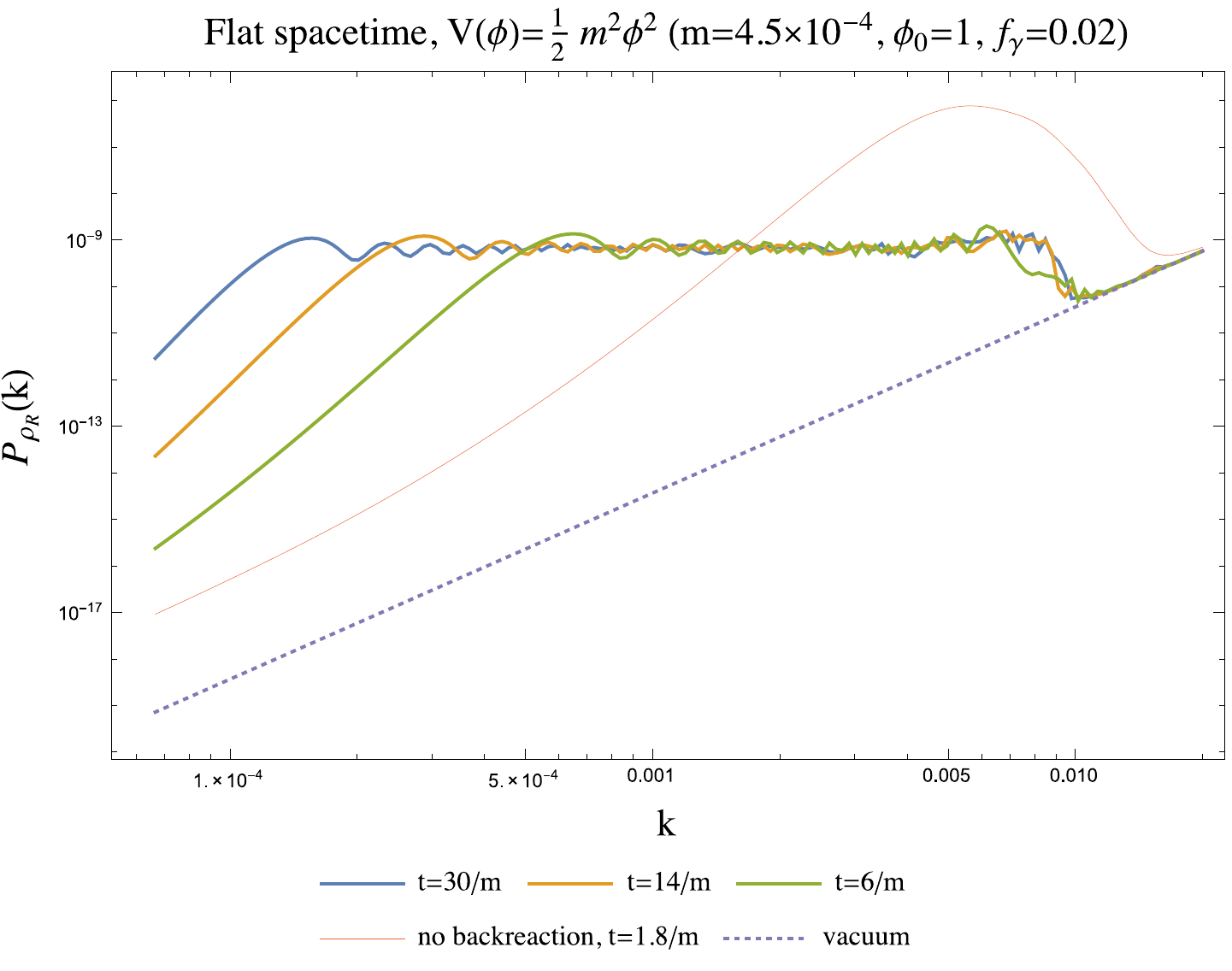}
    \caption{We show in the left panel the backreaction spectrum $P_{\cal B}(k)$,  defined in eq.~(\ref{spectrum}), at three different times during the evolution, using a quadratic potential and in flat spacetime. In the right panel we show  the spectrum of electromagnetic energy density $P_{\rho_R}(k)$, and also the vacuum energy density, which should be subtracted by renormalization. For both cases the flat region of the spectrum becomes wider as time goes on, since more infrared modes have enough time to get excited. The thin pink line represents the growth of modes in absence of backreaction, at a given time.}
    \label{PBflat}
\end{figure}

We see in fig.~\ref{Atflat} that the modes evolve in time exponentially in the initial stage of the evolution, until the backreaction term becomes important and starting from this time the modes grow at a very low rate. However as time goes on there are more modes, with lower values of $k$, which follow the same behavior. This leads to a spectrum $P_{\cal B}(k)$ which is almost flat in a band, which becomes wider and wider as time goes on, as can be seen in fig.~\ref{PBflat}.

The other important physical point, as we mentioned before, is that our coupling is an effective one, which we can trust only at $k<f_{\gamma}$ and  so 
we always cutoff the integral at $k=f_{\gamma}$. We did not attempt here at studying what happens in a more complete theory at higher energies, but we will comment on this in section~\ref{discussion}.
Note however that if the range of tachyonic modes ({\it i.e.} the ones for which $k<k_p\equiv |\dot{\phi}/f_{\gamma} |$) is always smaller than the cutoff $k_c\equiv f_{\gamma}$, then we do not need to consider such a high-energy completion of the theory. We will check  in the following in which cases we respect this condition.

\begin{figure}[t]
    \includegraphics[width=0.48 \textwidth]{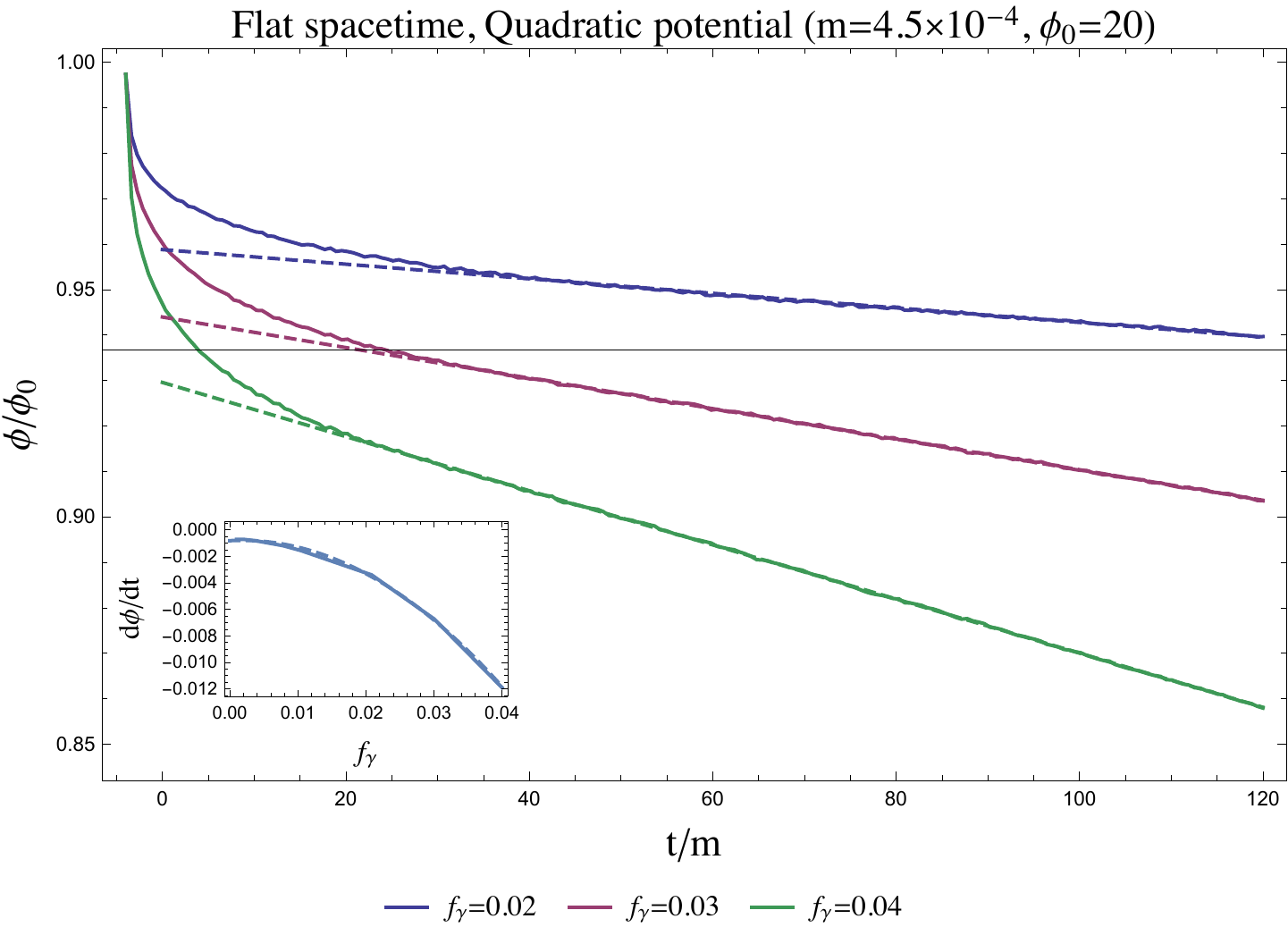}
      \includegraphics[width=0.48 \textwidth]{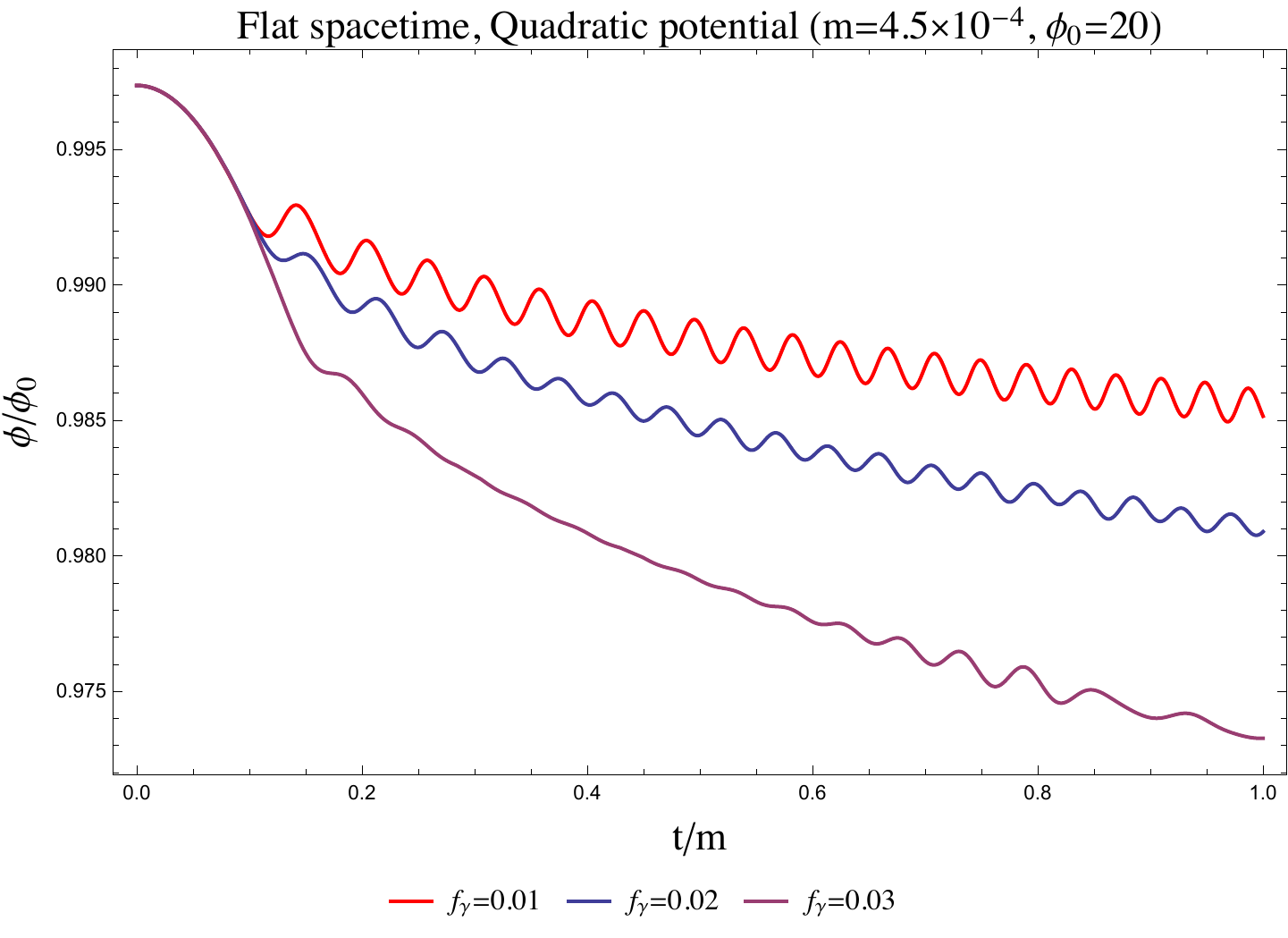}
    \caption{We show the time evolution of $\phi$ in the flat spacetime case, for a quadratic potential $V=\frac{1}{2}m^2 \phi^2$, for several values of $f_\gamma$. The dashed lines represent linear interpolations of the average asymptotic evolution. We also show, in the small panel, the dependence of such asymptotic slope $d\phi/dt$ on $f_\gamma$, which turns out to be quadratic. In the right plot we show the $\phi$ evolution for a smaller range of time, in order to show that there are small  oscillations at high frequency on top the average behaviors.}
    \label{fig:phiflat}
\end{figure}

\begin{figure}[t]
    \includegraphics [width=0.5 \textwidth]{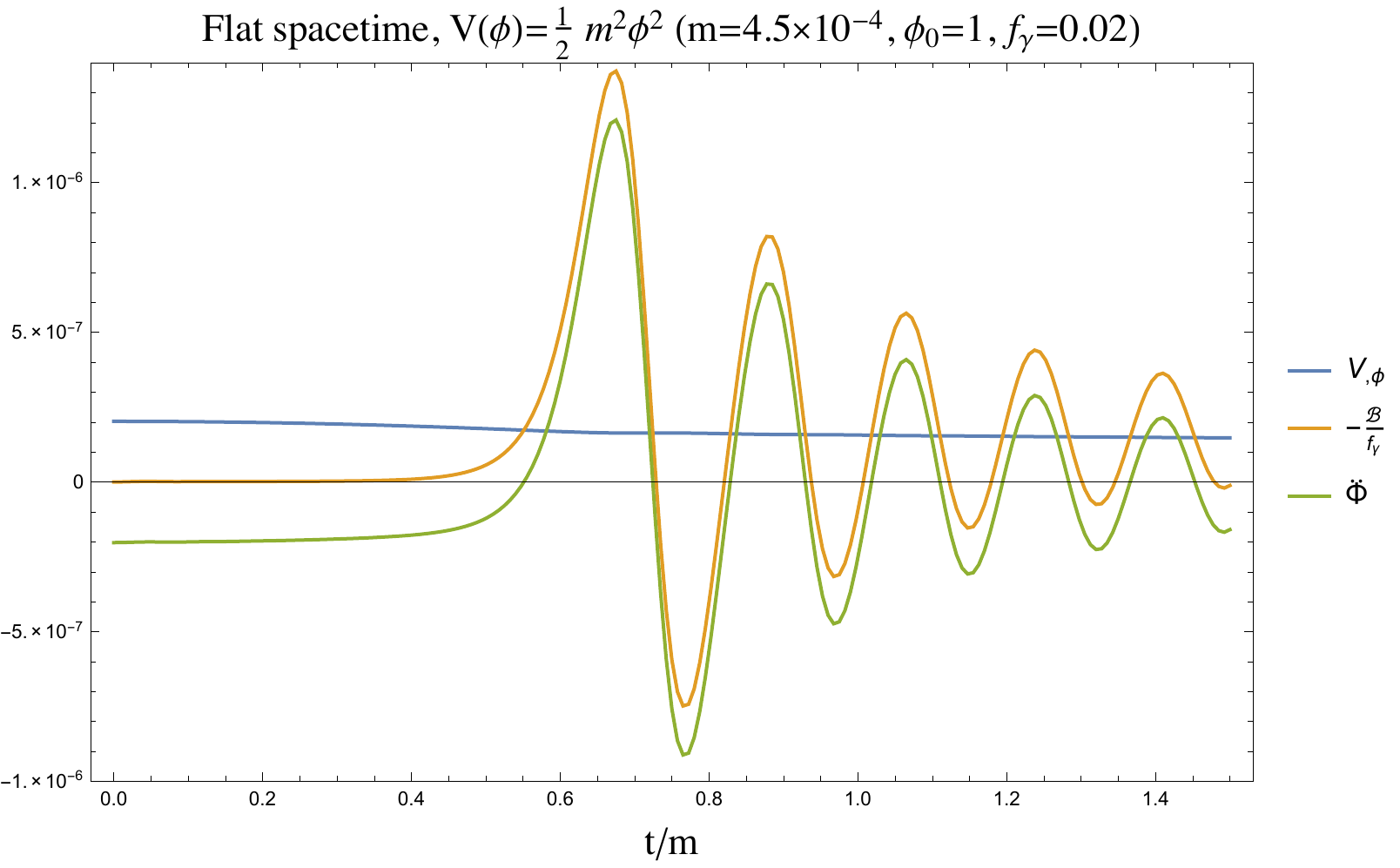}
       \includegraphics [width=0.51 \textwidth]{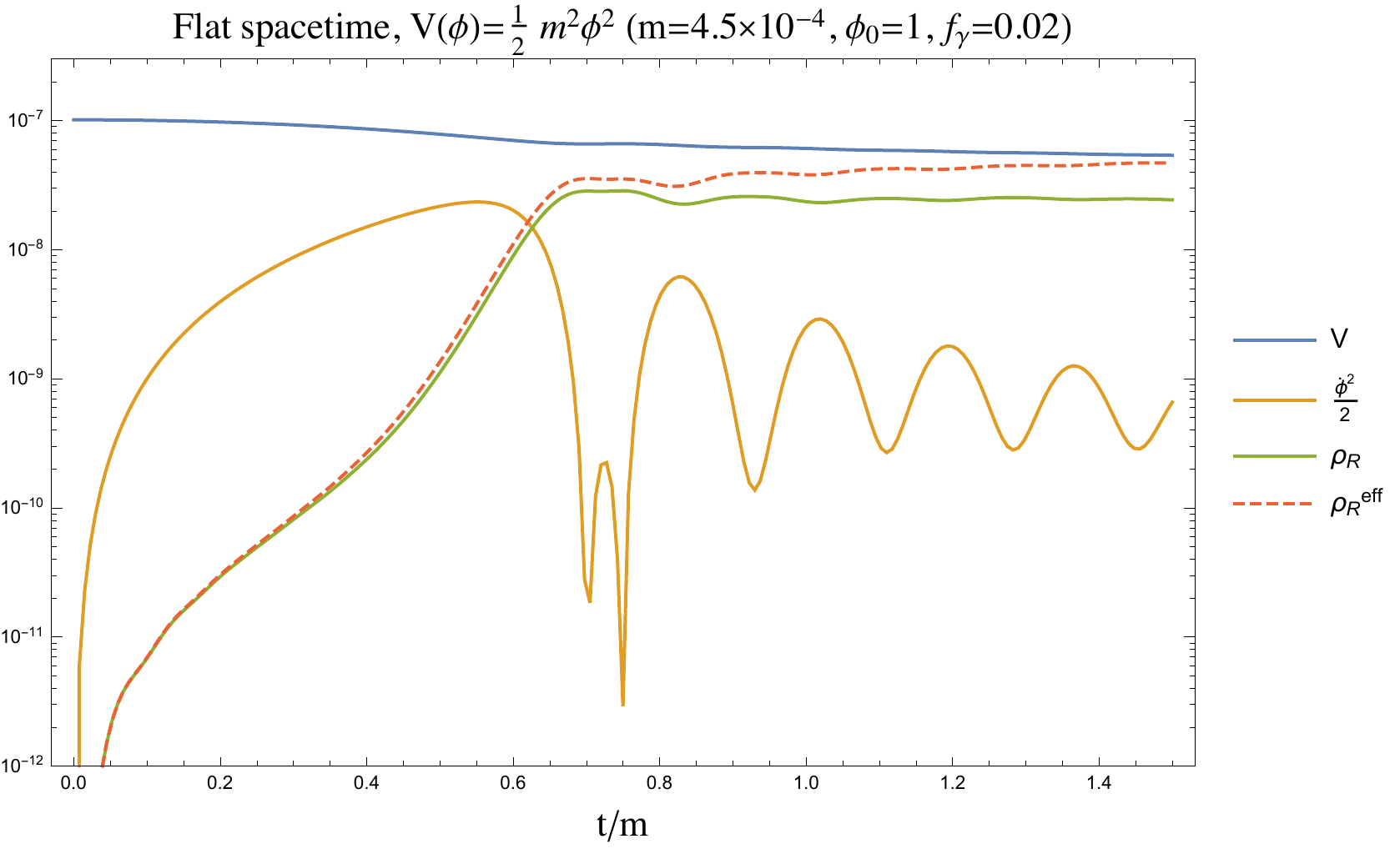}
    \caption{In the left panel we show the various terms (potential term, backreaction and $\ddot{\phi}$) in the equation of motion eq.~(\ref{phiflat}), for a quadratic potential and in flat spacetime. In the right panel we show the size of the various energy densities: potential and kinetic energy of $\phi$ and the renormalized $\rho_{R}$, computed using eq.~(\ref{RHOR}). The dashed line represents instead the result obtained using eq.~(\ref{rhoReqflat}), which agrees with the previous line at the beginning, but starts deviating at large time.}
    \label{Vphiplot}
\end{figure}

As anticipated we find that the evolution depends on the above discussed condition on $f_{\gamma}/\phi_0$. If we are in the regime $f_{\gamma}\gtrsim\phi_0$ the backreaction does not have enough time to become large, so nothing relevant happens  and the fields falls down normally.
If instead $f_{\gamma}$ is at least about one order of magnitude smaller than $\phi_0$ then the backreaction kicks in and we see a dramatic change in the behavior: the field slows down and ${\cal B}$ balances the potential for a very long time. Such a balance does not lead to a smooth behavior, but to rapid oscillations around an average slow evolution. Such average slow evolution leads eventually to an almost constant $\dot{\phi}$ regime, as we show in fig.~\ref{fig:phiflat} for several values of $f_\gamma$. In fig.~\ref{Vphiplot} we show the size of the various terms (potential term, backreaction and $\ddot{\phi}$)  in the equation of motion.
The dependence of $\dot{\phi}$ (its average behavior, disregarding oscillations) on $f_\gamma$ is found to be quadratic $\dot{\phi} = \alpha f^2_\gamma$, see fig.~\ref{fig:phiflat}, where $\alpha\approx -0.75$ is almost independent on the potential. 

A similar dynamics with two stages of evolution must take place also in an expanding background, as long as the time scales of the potential and of the evolution of the gauge fields are much smaller than the Hubble time $H^{-1}$ and as long as we follow the system for a time shorter than the Hubble scale ({\it i.e.} for much less than one efold of expansion of the metric).
In fact with a steep potential and a rapidly growing gauge field, we should first have a free fall regime just due to the potential and then the backreaction should turn on exactly in the same way as in flat spacetime and start dominating the evolution.  This should now lead to slow-roll and so to inflationary expansion of the metric. Some differences will appear when following the evolution for more than one or two efolds: (1) we have to take into account of the redshift of the  gauge fields modes, as inflation goes on for a long time, (2) there is a Hubble friction term also in the $\phi$ evolution. Such effects will lead to a third stage of the evolution, as we will address in the next section.

Finally we can also compute the energy density in the gauge field as:
\begin{eqnarray}
 \rho_R \equiv \left\langle \frac{B^2}{2} + \frac{E^2}{2} \right\rangle -\rho_R^{\rm vac} = \frac{1}{2} \sum_{h=\pm} \int \frac{d^3k}{(2\pi)^3}  \, \left[ \left( |\dot{A}_h|^2 +k^2  |A_{h}|^2 \right)-k \right] \label{RHOR}
\, , \end{eqnarray}
where $E$ and $B$ are electric and magnetic fields and where we have subtracted the vacuum contribution $\rho_R^{\rm vac}$, using $A_k=e^{i k t}/\sqrt{2 k}$, which takes care of the UV behavior. We also show its spectrum $P_{\rho_R}$, defined as $\rho_R=\int \frac{dk}{k} P_{\rho_R}$, in fig.~\ref{PBflat}. 
Another way to estimate $\rho_R$ is to rewrite eq.~(\ref{phiflat}) as a conservation equation:
\begin{eqnarray}
\dot{\rho}_\phi +  \frac{ {\cal B} \dot{\phi}}{f_\gamma} =0 \, , \qquad \rho_\phi \equiv V+\frac{\dot{\phi}^2}{2}
\, , \end{eqnarray}
which shows that there is energy dissipation in the scalar field. Since the total energy density is  conserved, such an energy flow must go into a gauge field energy density $\rho_{R}$ as follows:
\begin{eqnarray}
\dot{\rho}_{R} -  \frac{ {\cal B} \dot{\phi}}{f_\gamma}  =0  \label{rhoReqflat}
\, , \end{eqnarray}
By integrating numerically this equation the $\rho_{R}$ computed in this ``effective'' way should coincide with the integral over the modes given in  eq.~(\ref{RHOR}). We find that indeed they coincide in the initial stages of evolution, see fig.~\ref{Vphiplot}, but they start differing after some time, for unknown reasons. Note however that the two ways of computing $\rho_R$ nicely coincide instead in the full case in an FLRW background, discussed in the next section. Another important remark is that we can see in the same plot that the kinetic energy of $\phi$ is subdominant with respect to $\rho_R$.
%


\section{Inflation}\label{Inflation}

We consider now our system in a FLRW metric with conformal time $\tau$ 
\begin{eqnarray}
ds^2=a(\tau)(d\tau^2-d\vec{x}^2) \, .
\end{eqnarray}
The equations of motion are~\cite{Anber:2009ua}:
\begin{eqnarray}
&& \phi''+2 a H \phi'+ a^2 V_{,\phi}(\phi) + a^2 g_{,\phi}(\phi) {\cal B} =0 \, ,
\nonumber \\
&& A''_\pm+ (k^2\mp k g') A_\pm=0 \, ,
\nonumber \\
&& {\cal B} = \frac{\langle F\tilde{F}\rangle}{4} = \frac{1}{2 a^4} \int \frac{d^3 k}{(2\pi)^3} \, k \frac{d \left[ |A_+|^2 - |A_{-}|^2 \right] }{d\tau} \, ,
\label{coupled}
\end{eqnarray}
where a $'$ means a derivative with respect to $\tau$. Note that the gauge field equation is formally unchanged, while the scalar field has Hubble damping.
Note also that $\langle F\tilde{F}\rangle$ has now a dilution factor $a^{-4}$. In an inflationary background this may seem to induce a large suppression,  but it is actually not an issue: in fact such a dilution is counterbalanced by the explosive growth of the modes, keeping the value of $\langle F\tilde{F}\rangle$ roughly constant for a large number of efolds. The equation of motion for the scale factor is simply the Friedmann equation
\begin{eqnarray}
H^2\equiv \left( \frac{a'}{a^2} \right)^2 = \frac{V(\phi)+\frac{\phi'^2}{2 a^2}+\rho_R }{3 M^2_{Pl}} \, , \,\,\,\, \,\,
\rho_R \equiv \frac{1}{2 a^4} \sum_{\pm} \int \frac{d^3 k}{(2\pi)^3} \, \left[ |{A'}_h|^2 +k^2  |A_{h}|^2 -k \right]  
\, , 
\label{FLRW}
\end{eqnarray}
where $M_{Pl}=2.43 \times 10^{18}$ GeV is the reduced Planck mass and where the radiation energy density $\rho_R$ has been renormalized. Both $\rho_R$ and the field kinetic energy are subdominant compared to $V(\phi)$ if there is successful inflation.

Note  that  the $k$ modes are getting redshifted: as inflation goes on the long wavelength modes go superhorizon and become less relevant in $\langle F\tilde{F}\rangle$ (because of the weight $k \, d^3k$), but at the same time new shorter wavelength growing modes are constantly generated. The expected outcome is an almost stationary process with an almost constant backreaction, which can be checked by integrating the full system of equations. Moreover, considering the full system also includes the intermediate case in which $\phi$  could have both electromagnetic and gravitational friction, in comparable amounts.

As we said, we expect that the initial behavior of $\phi$ must be very similar to the flat spacetime case, at least when following the system for less than about one efold. We are interested in fact in a case in which the potential is steep and thus its typical timescale is much smaller than $H^{-1}$; the same is true for the gauge fields, whose evolution is even faster. So the onset of backreaction  happens under the same conditions that we have seen in the flat spacetime case, namely that $ \phi_0 \gg f_\gamma$. However, once the backreaction sets in, it leads to a very slow evolution of $\phi$ and so the Hubble expansion  may start to become important. This is indeed the case, as we discuss below.

%

Such a stage of evolution in an inflationary background has been studied in~\cite{Anber:2009ua}, where an analytical  estimate of ${\cal B}$ has been given. Assuming that backreaction can indeed become large the authors have worked under the hypothesis that a regime with constant $\dot{\phi}$ can be reached. In such a background they considered a de Sitter metric and computed analytically the evolution of the gauge fields setting an initial condition at infinite past time and looking for the asymptotic solution at future infinity. We have checked that  their approximate solution
\begin{eqnarray}
A^{an}_\pm=\frac{1}{\sqrt{2k}} \left( \frac{k}{2\xi a H}\right)^{1/4} e^{\pi \xi-2\sqrt{2\xi k/(a H)}} \, ,
\end{eqnarray}
gives the correct size for both the imaginary and the real part of the full numerical solution, as a late time behavior, assuming constant $\dot{\phi}$ and $H$, where $\xi\equiv \frac{\phi_{,N}}{2 f_\gamma}$. 
After an integration in $k$ they have then derived the following equation 
for $\dot{\phi}$
\begin{eqnarray}
 \ddot{\phi}+3 H \dot{\phi}+V_{,\phi} = {\rm sign}({\xi}) \frac{{\cal I}}{f_\gamma} \left( \frac{H}{\xi}\right)^4 e^{2 \pi |\xi|} \, , \label{eqsorbo} 
 \end{eqnarray}
where ${\cal I}=2.4 \times 10^{-4}$ (we have  introduced the $ {\rm sign}(\xi)$  to include also the case $\xi<0$). Given this equation one could infer the value of $\xi$ just equating approximately $V_{,\phi}$ to the backreaction term, as suggested in~\cite{Anber:2009ua}. A caveat is that this would lead to two solutions: one at very small $\xi$ and the other one at $\xi$ larger than ${\cal O}(1)$. We checked  that solving numerically eq.~(\ref{eqsorbo}) leads rapidly to the  first solution or the second solution, depending on the initial velocity of the field. Such ambiguity is not present in our full numerical solutions, which we discuss in the next section, and the solution that is reached after 1-2 efolds agrees qualitatively well with the second solution, as discussed in the next section. Such a second solution is approximately given by eq.~(11) of~\cite{Anber:2009ua}:
 \begin{eqnarray}
\xi \approx \frac{1}{2\pi} \log \left(\frac{9 M_{Pl}^4 f_\gamma V_{, \phi}}{{\cal I} V^2} \right) \, . \label{eqxisorbo}
 \end{eqnarray}

\section{Potentials and Results} \label{results}

In order to integrate the equations in a more efficient way we change variable from $\tau$ to the number of efolds $N\equiv \log(a)$, choosing as initial value $N_0=0$ and using eqs.~(\ref{initial}) as initial conditions (but replacing $d/dt$ with $d/d\tau$), leading to
\begin{eqnarray}
&& H^2 \phi_{,NN}+3 H^2 \phi_{,N} +H H_N \phi_N+ V_{,\phi}(\phi) +  g_{,\phi}(\phi) {\cal B} =0 \, ,
\nonumber \\
&& e^{2N}H^2 (A_{\pm, NN}+ A_{\pm, N})+ e^{2N}H H_{,N} A_{\pm, N} + (k^2\mp  k e^N H g_{,N}) A_{\pm}=0  \, ,
\nonumber \\
&& {\cal B} = \frac{e^{-4 N}}{2} \int \frac{d^3 k}{(2\pi)^3} \, k H e^N \frac{d \left[ |A_+|^2 - |A_{-}|^2 \right] }{dN} \, , \nonumber \\
&& H^2 =\frac{V+\rho_R}{3 M_{Pl}^2-\phi_N^2/2} \, .
\label{coupled} \end{eqnarray}
We analyze potentials of the following types 
\begin{eqnarray}
(i) \qquad V(\phi) &=& \frac{1}{2} m^2 \phi^2   \, \\
(ii) \qquad V(\phi) &=& \Lambda^4 \left[1-\cos\left(\frac{\phi}{M} \right) \right] \, . \label{Axion}
 \end{eqnarray}
The first is a simple minimal choice, while the second  is particularly interesting because it is an axion potential (plus a constant, such that the minimum at $\phi=0$ has $V(0)=0$). Let us in fact stress a crucial point here, taking as an illustration the QCD axion, though keeping in mind that this is generic for any axion-like field. In the QCD case the coupling which is needed to solve the strong CP problem is the one with gluons
\begin{eqnarray}
{\cal L}_{G}= \frac{\phi}{f_G} G_{\mu \nu} \tilde{G}^{\mu \nu}   \, .
 \end{eqnarray}
 Such a coupling to non-abelian fields breaks the shift symmetry  $\phi \rightarrow \phi+ c $, leaving only the symmetry $\phi \rightarrow \phi+ 2\pi f_G $ and inducing a periodic potential of the above form with $\Lambda=\Lambda_{QCD}$ and $M=f_G$
(note that this is only an approximation for the QCD axion, which is good around minima~\cite{DiVecchia:1980yfw}).
The relevant point here is that we may assume that the field has a starting point in a generic position, displaced from the minimum by an amount of order $\phi_0 \approx f_G$. 
 Therefore, as we have discussed, the condition we need for inflation is that $f_\gamma\ll f_G$, by about one or two orders of magnitude. Note that, although in vast majority of analyses of the QCD axion  such two couplings are considered to be comparable, they are actually two independent parameters. Moreover even if the coupling to photons is much larger it does not break the $\phi$ shift symmetry and theferore it cannot induce large loop corrections to the gluon coupling. It is therefore allowed to have $1/f_\gamma\gg 1/f_G$.

Thus, a crucial observation is that an axion field coupled stronger to abelian than non-abelian gauge fields is a viable candidate for dissipative inflation. It is very interesting to check whether using the QCD axion itself for inflation with such couplings can work, which we discuss later.


It turns out to be possible to integrate eqs.~(\ref{coupled}), with a few difficulties. First, as seen also in the previous section, one needs to include  a large enough density of modes so that the evolution converges to a smooth one. Second, one needs to take care of the fact that a huge range of $k$'s has to be used during inflation, since the relevant ones are constantly redshifted away. Moreover, as we have discussed, we do not have control on the theory when physical momenta are above the cutoff. However, unless $f_\gamma$ is extremely small, typically the modes above the cutoff  would just anyway oscillate.  We deal therefore with this situation by freezing the modes in the numerical solution, as long as they satisfy $k/a>f_{\gamma}$  and then starting integrating only when such a condition is met. However, when $f_\gamma$ is very small the modes with $k/a>f_{\gamma}$ could also be in the unstable region: in this case the effective theory is not adequate and the full answer should come from a UV complete theory.

\begin{figure}[t]
    \centering
    \includegraphics[width=0.6 \textwidth]{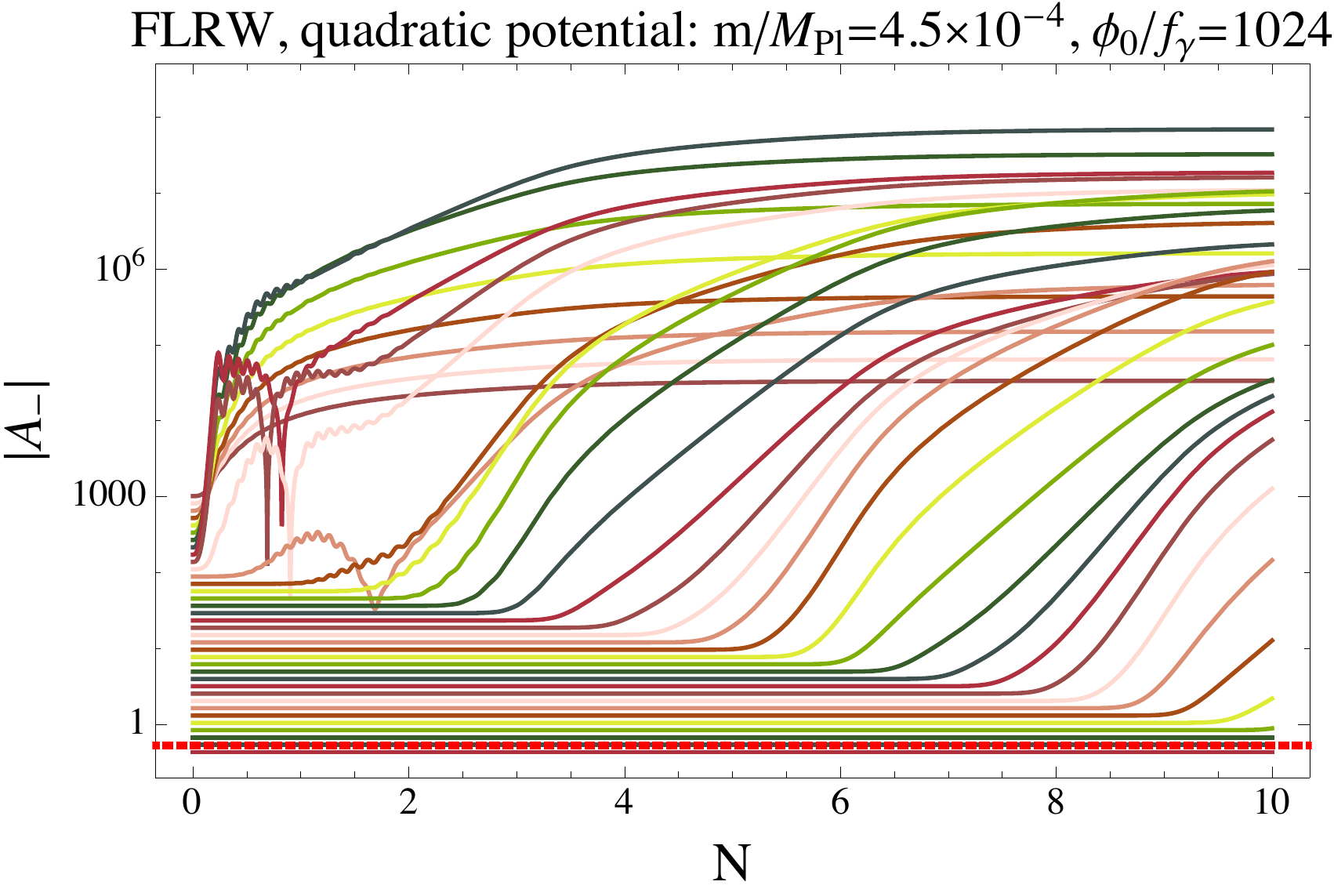}
    \caption{We show the evolution of the the gauge field $A_-$ as a function of the number of efolds, ranging from $N_i=0$ up to $N_f=11$, for different $k$'s  in a FLRW inflating background. The last line in the bottom of the plot corresponds to a mode with maximal value of the momentum $k_{max}=a_f f_\gamma$ while the other lines correspond to momenta which decrease with equal logarithmic spacing down to the minimal value of  $k_{min}=a_i f_\gamma/50$, where $a_i=1$ and $a_f=e^{N_f}$ are respectively the initial and final value of the scale factor. We used here units of $\phi_0=M_{Pl}=10$.}
     \label{Atinfl}
\end{figure}

\begin{figure}[t]
    \centering
    \includegraphics[width=0.46 \textwidth]{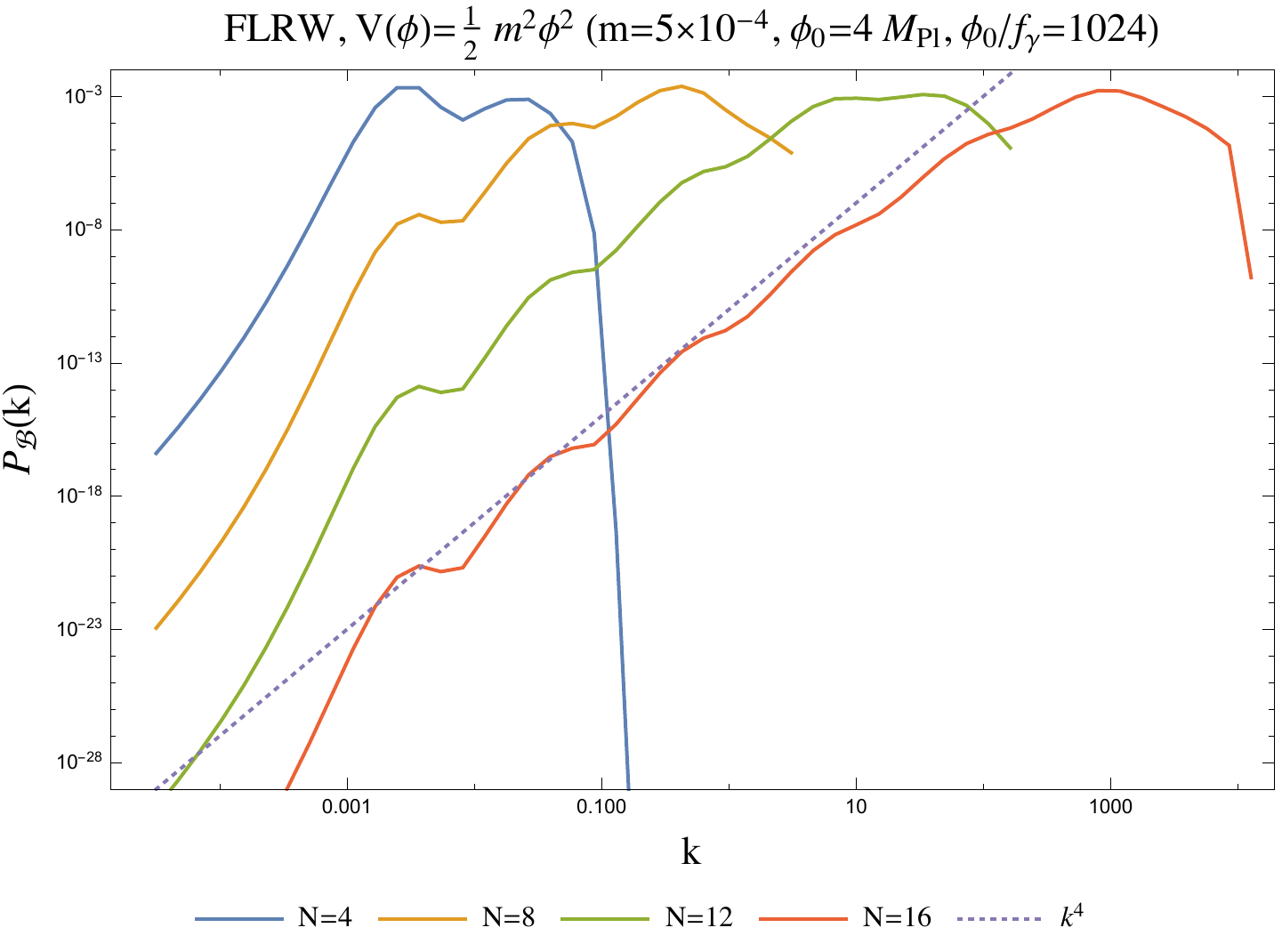}
    \includegraphics[width=0.46 \textwidth]{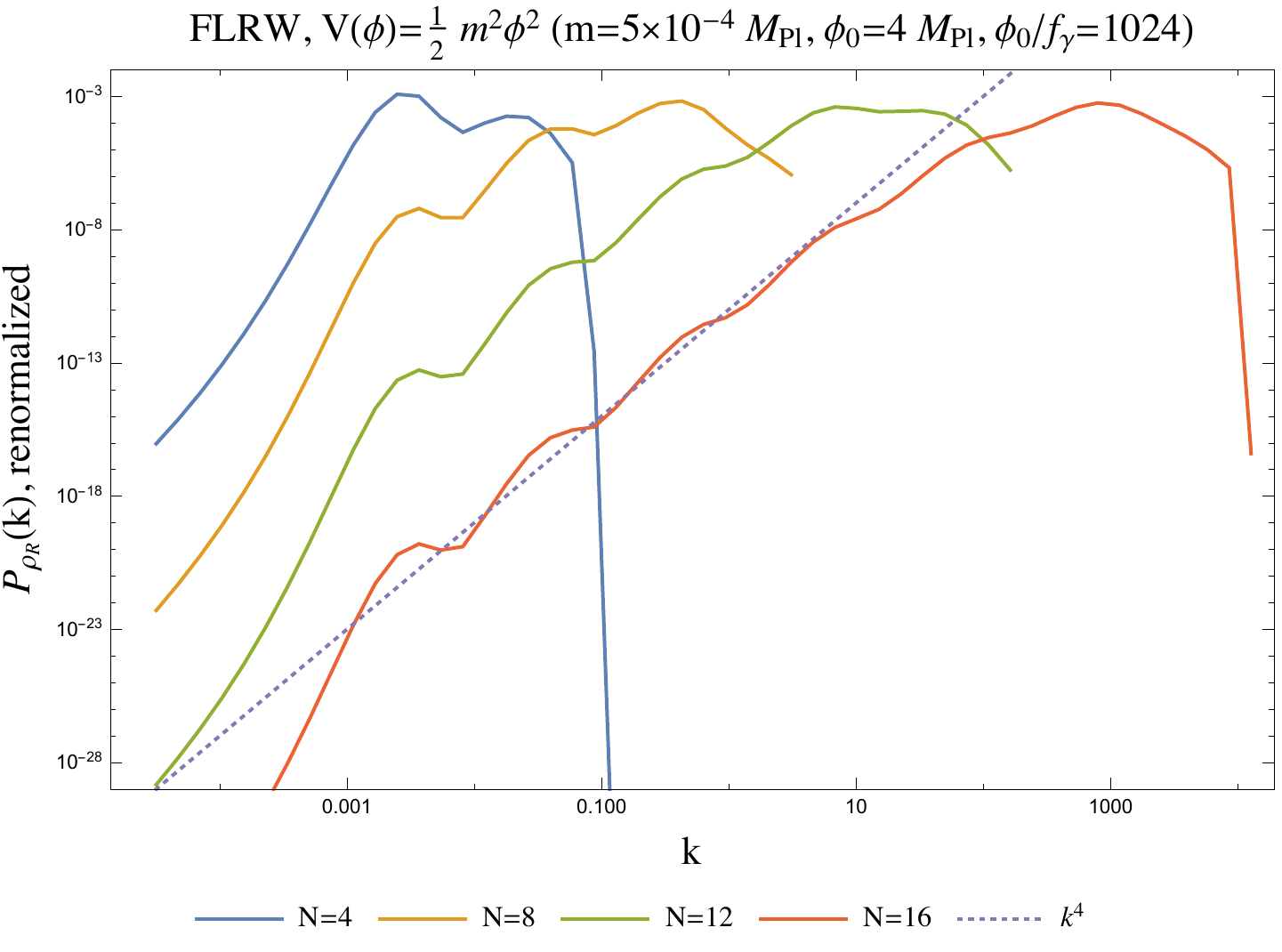}
    \caption{We show in the left panel the backreaction spectrum $P_{\cal B}(k)$,  defined in eq.~(\ref{spectrum}), at three different numbers of efolds $N$ during the evolution, using a quadratic potential and in a FLRW inflating background. In the right panel we also show  the spectrum of the renormalized electromagnetic energy density $P_{\rho_R}(k)$.  We show as a reference, in dotted line, a $k^4$ behavior. Both plots have a range of modes between $k_{min}=a_i f_\gamma/50$ and  $k_{max}=a_f f_\gamma$, where $a_i$ and $a_f$ are respectively the initial and final value of the scale factor.}
    \label{PBFLRW}
\end{figure}

\begin{figure}[t]
    \includegraphics[width=0.5 \textwidth]{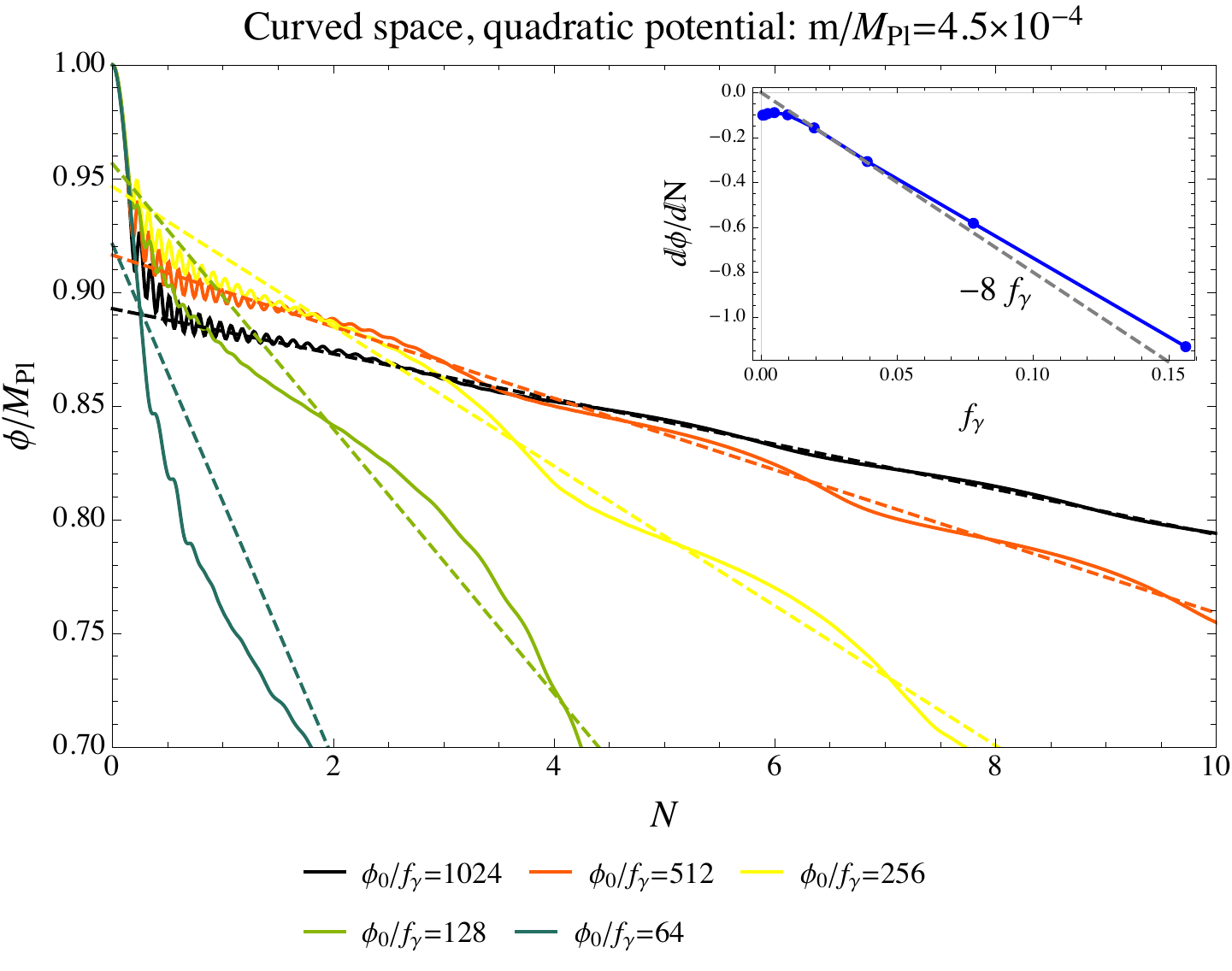}
    \includegraphics[width=0.495 \textwidth]{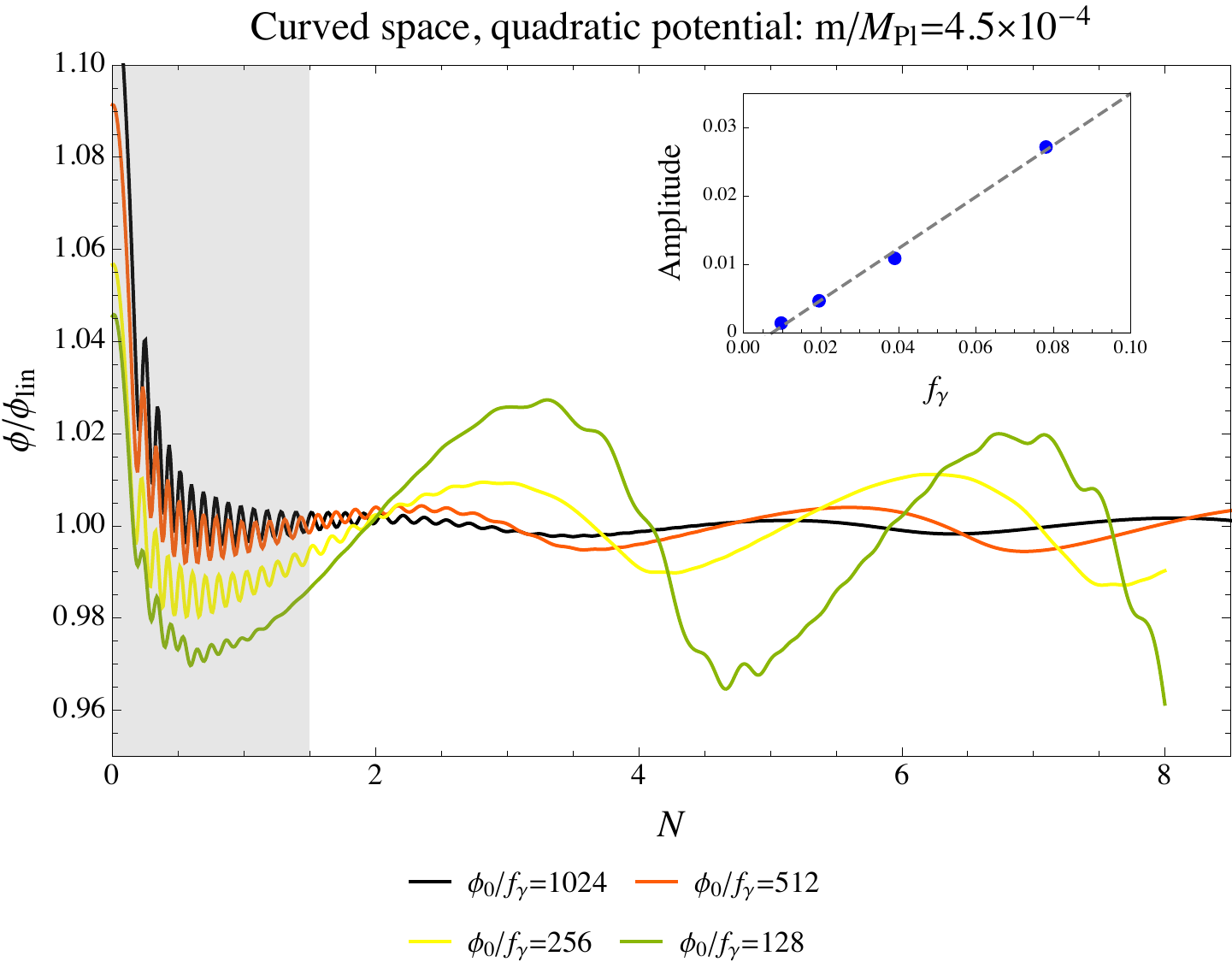}
    \caption{We show in the left panel the evolution of $\phi$ in a FLRW background, as a function of the number of efolds $N$, for several values of $f_\gamma$. We show also linear interpolations in dashed lines and in the small panel we display the scaling of their slopes with $f_\gamma$: note that $d\phi/dN$ goes roughly linearly as $8 f_\gamma/\phi_0$ (except at very small $f\gamma$, where modes of $k>f_\gamma$ start being important, and are ignored in our effective treatment). This agrees quite well with the analytical scaling $d\phi/dN\approx 2 \xi f_\gamma $ of~\cite{Anber:2009ua}, since $\xi\approx 3.5$ for these parameters. In the right panel we show the ratio of the full field evolution vs. the linear interpolation to highlight the presence of oscillations with a period of a few efolds, after an initial transient. We show in the small panel the scaling of the amplitude of such oscillations with $f_\gamma$.}
    \label{PlotphiFLRW}
\end{figure}

We show the mode evolution in fig.~\ref{Atinfl} and the relative spectra in fig.~\ref{PBFLRW}. The field evolution is shown in fig.~\ref{PlotphiFLRW}, for a quadratic potential and for different values of $f_\gamma$.
Some features are very striking. When the friction kicks in, it is very efficient and slows down dramatically the evolution of the field. Even if we integrate for only 10-20 efolds it is obvious by visual extrapolation that the slow roll keeps going easily for much more than 60 efolds, as long as $f_{\gamma}$ is smaller than a threshold value. Our numerical results agree qualitatively with~\cite{Cheng:2015oqa}, although in that work the parameters have been chosen so that most of the inflationary trajectory is due to usual Hubble friction dominated evolution and only a few more efolds due to gauge field backreaction. This has been motivated by the need of  keeping the parameter $\xi$ smaller than roughly 2, which supposedly is required to avoid large non-gaussianities. We do not apply such constraints, since as we argue here below, a reliable calculation of the curvature perturbations  has not been performed yet in the backreaction dominated case, and so this is even more the case for its three-point function.

\begin{figure}[t]
  \begin{tabular}{p{0.3\textwidth} p{0.3\textwidth} p{0.3\textwidth}}
 \vspace{0pt}  \includegraphics[width=0.3 \textwidth]{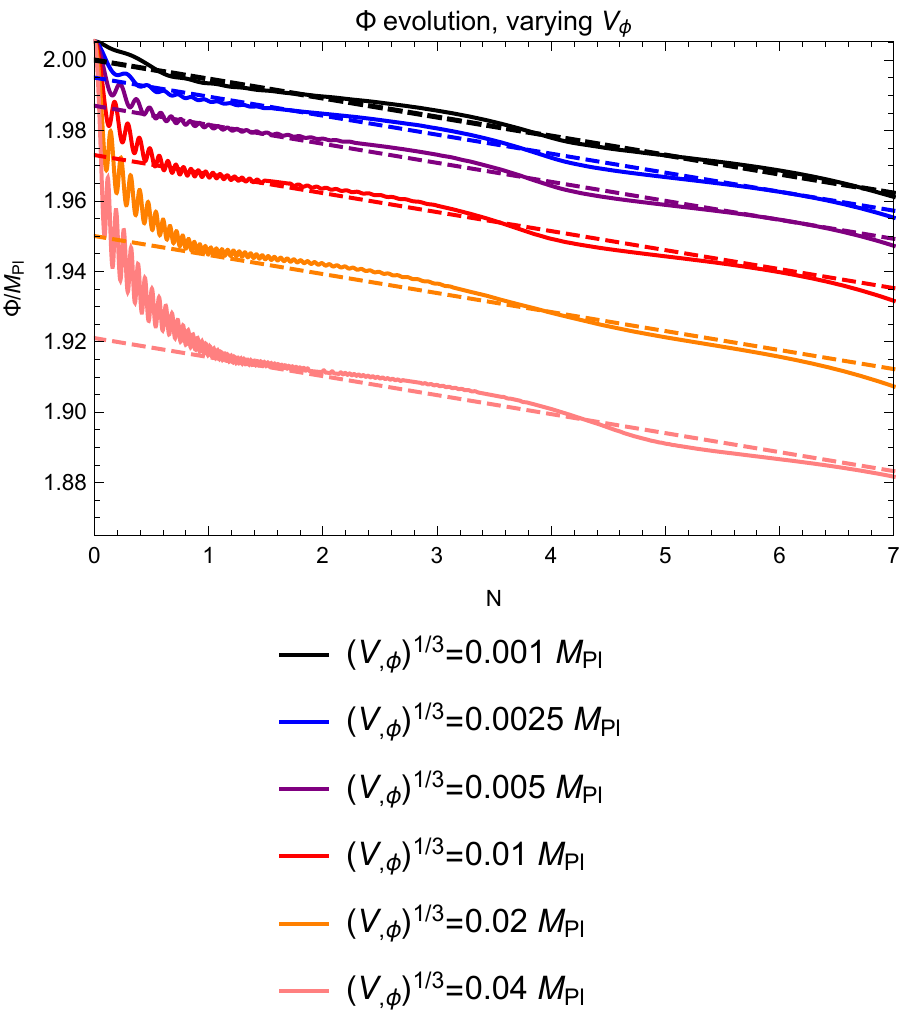} &
  \vspace{1pt}         \includegraphics[width=0.3 \textwidth]{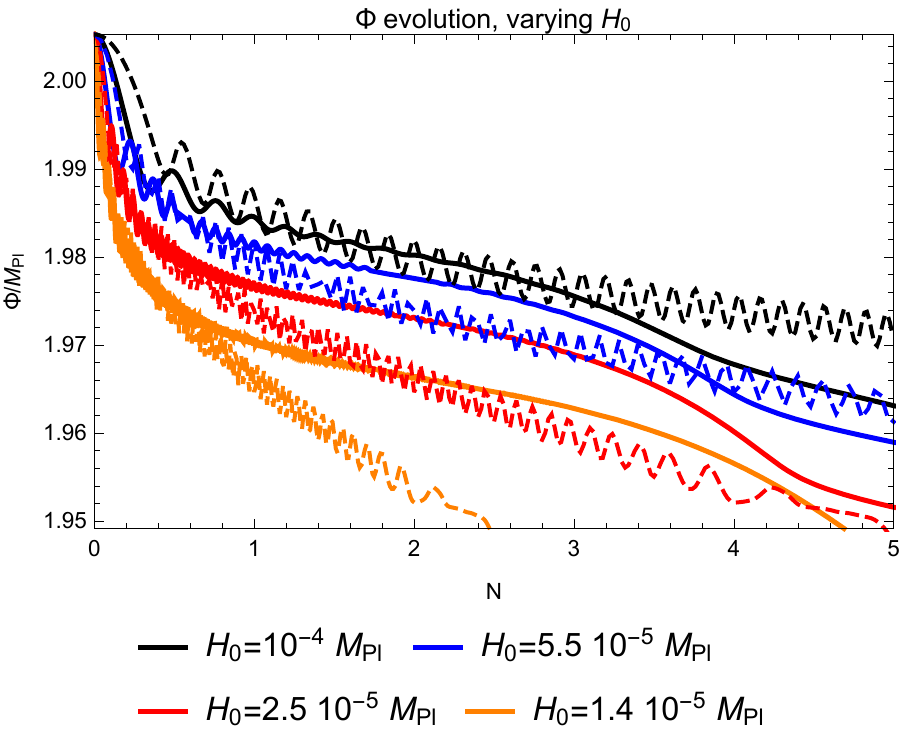} &
    \vspace{-1pt}        \includegraphics[width=0.3 \textwidth]{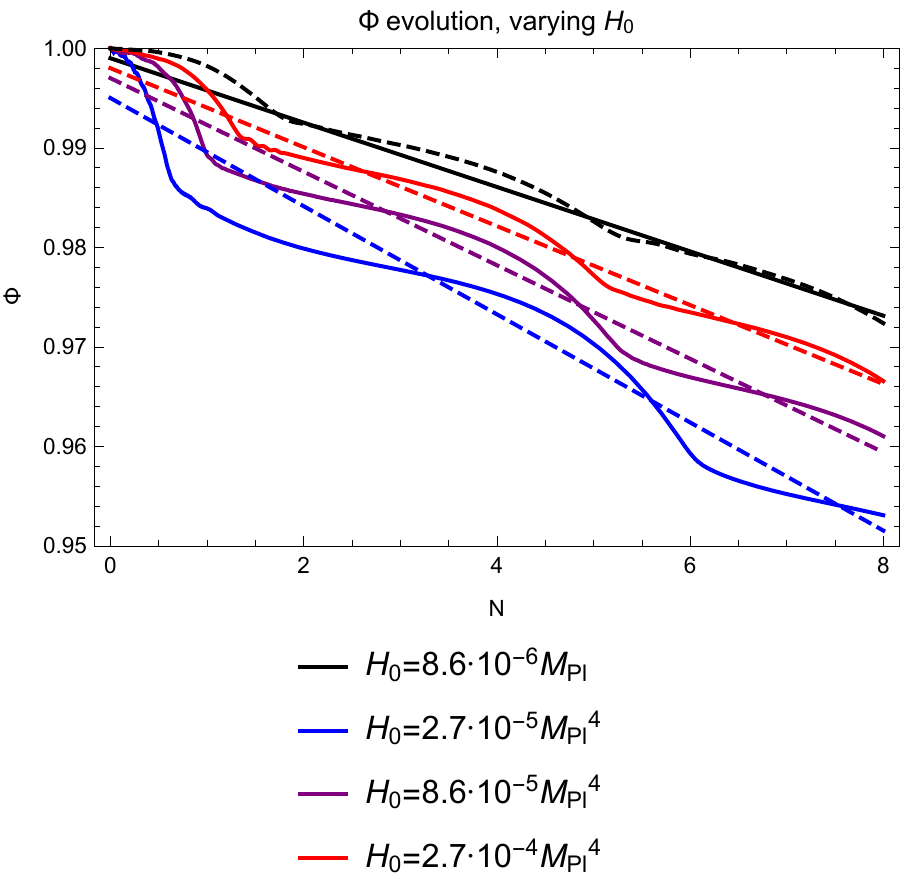} 
    \end{tabular}  
          \caption{In the left panel  we show the dependence of the field evolution on $V_{,\phi}$, keeping fixed all other parameters, including the height of the potential $V(\phi_0)$ (achieved by adding a suitable constant to it). Note that the initial transient stage at $N\lesssim1$, as in the flat spacetime case depends on the size of $V_{,\phi}$. Instead in the subsequent stage, a dependence on $V_{,\phi}$ is not visible. The dotted lines are linear interpolations of  the final stage, which all have the same slope of about $0.0027$, corresponding roughly to $740$ efolds.  We have chosen here a quadratic potential, with $H_0=5.5 \times 10^{-5} M_{Pl} $, $\phi_0=2 M_{Pl}$, $f_\gamma=\phi_0/2500$.
    In the central panel we show in solid lines the dependence of the field evolution on $H_0$, keeping fixed all other parameters, including  $V_{,\phi}$. Note that after the initial transient stage at $N\lesssim1$, there is a second stage, in which the slope $\phi_{,N}$ depends very weakly on $H_0$.  The dotted lines here would be the field evolutions in the flat spacetime case (using $N=H_0 t$ with $H_0^2=V(\phi_0)/(3 M^2_{Pl}$). We have chosen here a quadratic potential, with $m=6 \times 10^{-5} M_{Pl} $, $\phi_0=2 M_{Pl}$, $f_\gamma=\phi_0/2500$. In the right panel we considered a wider range for $H_0$, so that a weak dependence on it becomes visible. The dashed lines are the analytical expression given by eq.~\ref{eqxisorbo} (adjusted with a constant to account for the initial transient). Here $m=6.6 \times 10^{-6} M_{Pl}$ and $\phi_0=M_{Pl}$.}
    \label{HVdependence}
\end{figure}

The main results of our analysis are the following:
\begin{itemize}
\item If the field starts at rest it begins falling down rapidly but, if $\phi_0/f_{\gamma}\gtrsim {\cal O}(10)$, it is suddenly slowed down reaching a second stage of evolution similar to the one studied in the flat spacetime case, where its velocity $\dot{\phi}_1$ tends to a constant (plus some high frequency oscillations) and scales  linearly as a function of $f_\gamma$ as in fig.~\ref{fig:phiflat}. These two stages can be seen in the first efold of evolution, looking also at fig.~\ref{HVdependence}.
\item If this friction-dominated regime lasts for at least about one efold, the field enters a third stage in which the Hubble friction and the redshift of modes becomes important: high frequency oscillations disappear and $\phi$ reaches another asymptotic velocity $\dot{\phi}_2$. This stage is quite well approximated by $d\phi/dN\approx 2 \xi f_\gamma $, as in~\cite{Anber:2009ua} (and so $\dot{\phi}_2\approx 2 \xi f_\gamma H$). Therefore the total number of efolds is approximately $\phi_0/(2 \xi f_\gamma)$.
\item Another relevant feature is that the numerical solutions show a  stable evolution,  but with slow superimposed oscillations, as shown in the right panel of fig.~\ref{PlotphiFLRW}. Such oscillations do not seem  to disappear for any of the analyzed values of $f_\gamma$ and they do not vanish by increasing the precision of the numerical integration or the number of modes. Therefore we believe they are a physical feature of the evolution: the field starts falling down but then the backreaction stops it, as a consequence the gauge modes are afterwards less amplified and the field can move again until the same pattern repeats itself. Such a patterns is consistent with the spikes found also in~\cite{Cheng:2015oqa}. The  amplitude of such oscillations is found to depend linearly on $f_\gamma$ and it is smaller than about ${\cal O}(0.1\%)$ for cases with a total number of efolds of at least 60. The period instead seems almost independent on $f_\gamma$, and turns out to be around 4-5 efolds. This phenomenon must leads to potentially observable consequences, at the level of the density and tensor perturbations, most likely as oscillations in the power spectrum and therefore as a potential unique signature of the model. It is tempting to mention that the presence of such oscillations might be associated  to features of the CMB spectrum, which do not fit in the usual single field slow-roll scenario: for instance the power suppression on large scales~\cite{Ade:2015lrj} or a possible oscillation in the power spectrum~\cite{Ade:2015lrj,Meerburg:2013dla}. We postpone anyway a detailed study to future work, also because in the present paper we do not use any constraints from perturbations around such a background solution, as we will explain later. 
\item If we decrease the energy scale of the potential keeping fixed all other parameters, we decrease $H$ and so during the first friction-dominated stage  the field falls down very quickly since $d\phi/dN\simeq \dot{\phi}_1/H\simeq \alpha f^2_\gamma/H$. If the evolution happens at very low energy $H$ it can happen therefore that the field falls to zero in much less than 1 efold, and so in this case the second stage of friction dominated inflation can never actually start. This is a concern in cases such as inflation with the QCD axion, where $H$ is very small since the scale is $\Lambda_{\rm QCD}$.
\end{itemize}

%

\begin{figure}[t]
       \includegraphics [width=0.52 \textwidth]{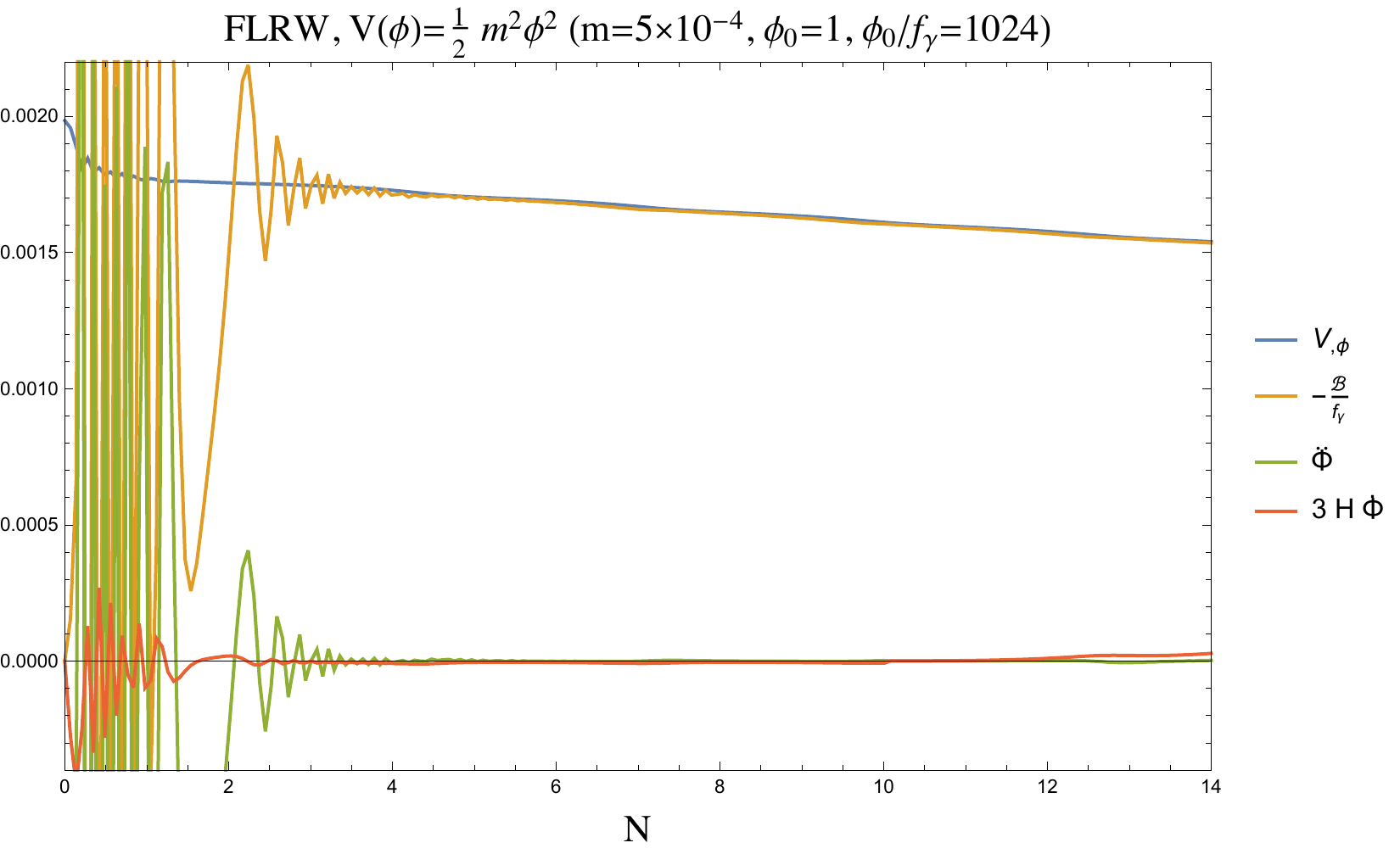}
    \includegraphics [width=0.524 \textwidth]{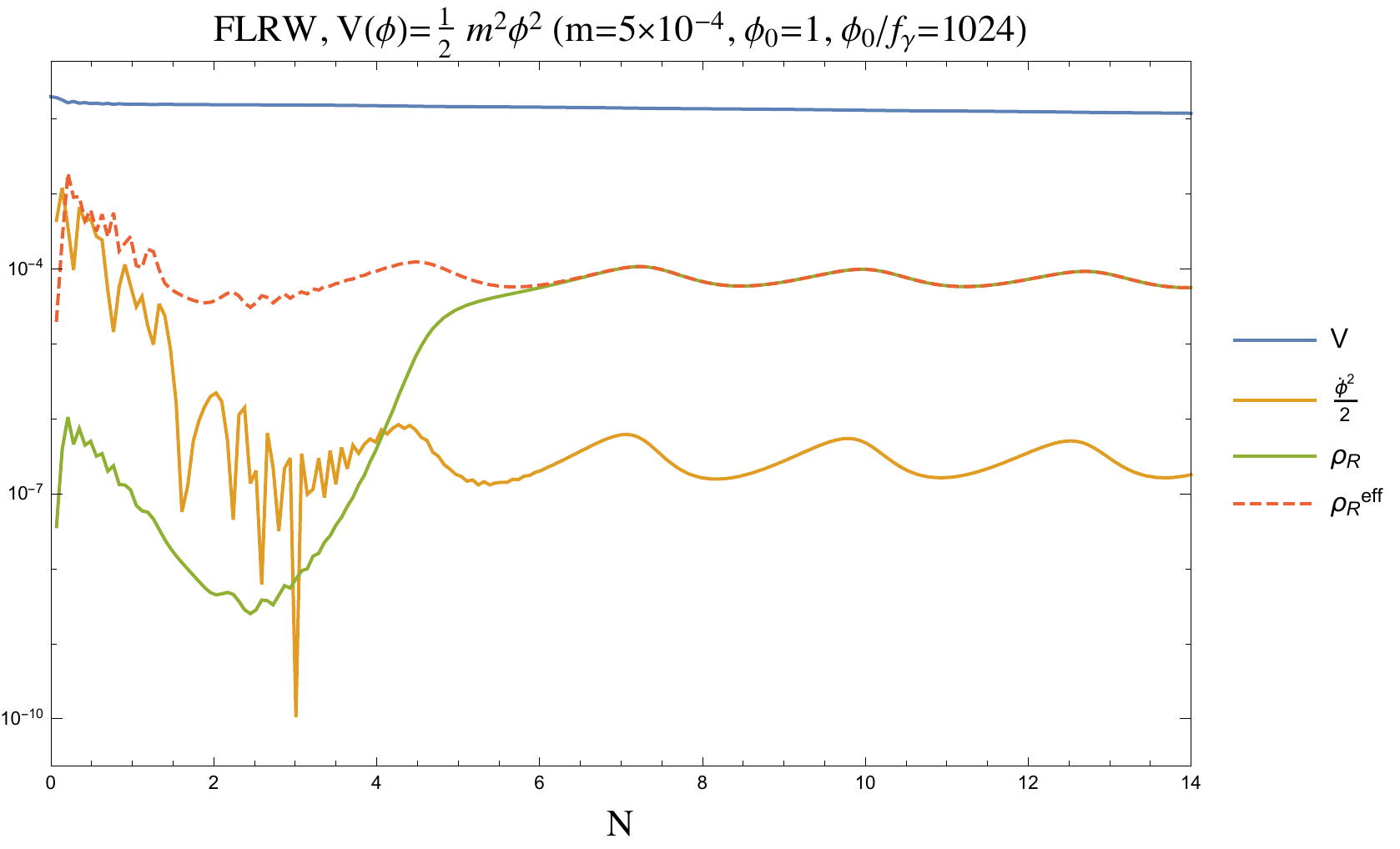}
        \caption{In the left panel we show the various terms (potential term, backreaction and $\ddot{\phi}$) in the full equation of motion eq.~(\ref{coupled}), for a quadratic potential and in FLRW spacetime. In the right panel we show the size of the various energy densities: potential and kinetic energy of $\phi$ and the renormalized $\rho_{R}$, computed using eq.~(\ref{FLRW}). The dashed line represents instead the effective way of computing the energy density, as in eq.(\ref{rhoReqinfl}). }
    \label{VphiplotFLRW}
\end{figure}

Then, we show the dependence of the asymptotic inflationary $\dot{\phi}_2$ on $V'$ and $V$ in fig.~\ref{HVdependence}, which confirm that the dependence is very weak, as expected from eq.~(\ref{eqxisorbo}). We also show in the left panel of fig.~\ref{VphiplotFLRW} the size of the various terms in the $\phi$ equation of motion.

Note, moreover, that the background equations can be rewritten as:
\begin{eqnarray}
&& 3 M_{Pl}^2 H^2 = \rho_\phi+\rho_R \, ,\\
&& \dot{\rho}_{\phi} + 3 H \dot{\phi}^2+\frac{{\cal B}}{f_\gamma} \dot{\phi}=0 \, , \\
&& \dot{\rho}_{R} + 4 H \rho_R-\frac{{\cal B}}{f_\gamma} \dot{\phi}=0  \, ,\label{rhoReqinfl}\\
 && \rho_\phi=V(\phi)+\dot{\phi}^2/2 \, .
\end{eqnarray}
As a response to the background $\phi$ evolution an electromagnetic field is produced, which on average is homogeneous and isotropic, and it turns out that  $\rho_R \gg \dot{\phi}^2/2$, as can be seen from the right panel of fig.~\ref{VphiplotFLRW}. This fact is crucial for the treatment of perturbations in such a scenario, as discussed below.

\begin{figure}
  \begin{tabular}{p{0.47\textwidth} p{0.47\textwidth}}
 \vspace{0pt} \includegraphics[width=0.47 \textwidth]{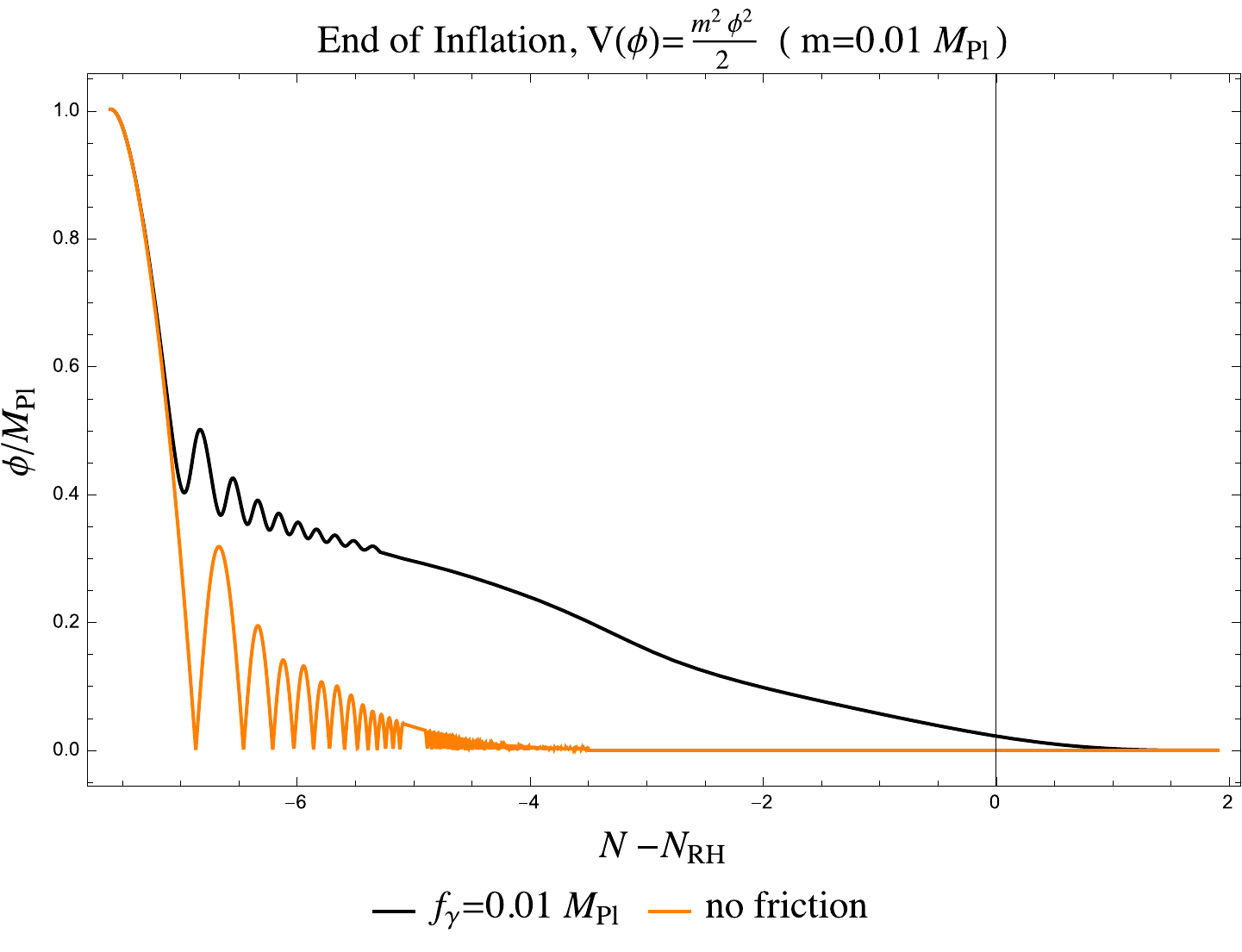} &
  \vspace{-7.2pt}   \includegraphics[width=0.503 \textwidth]{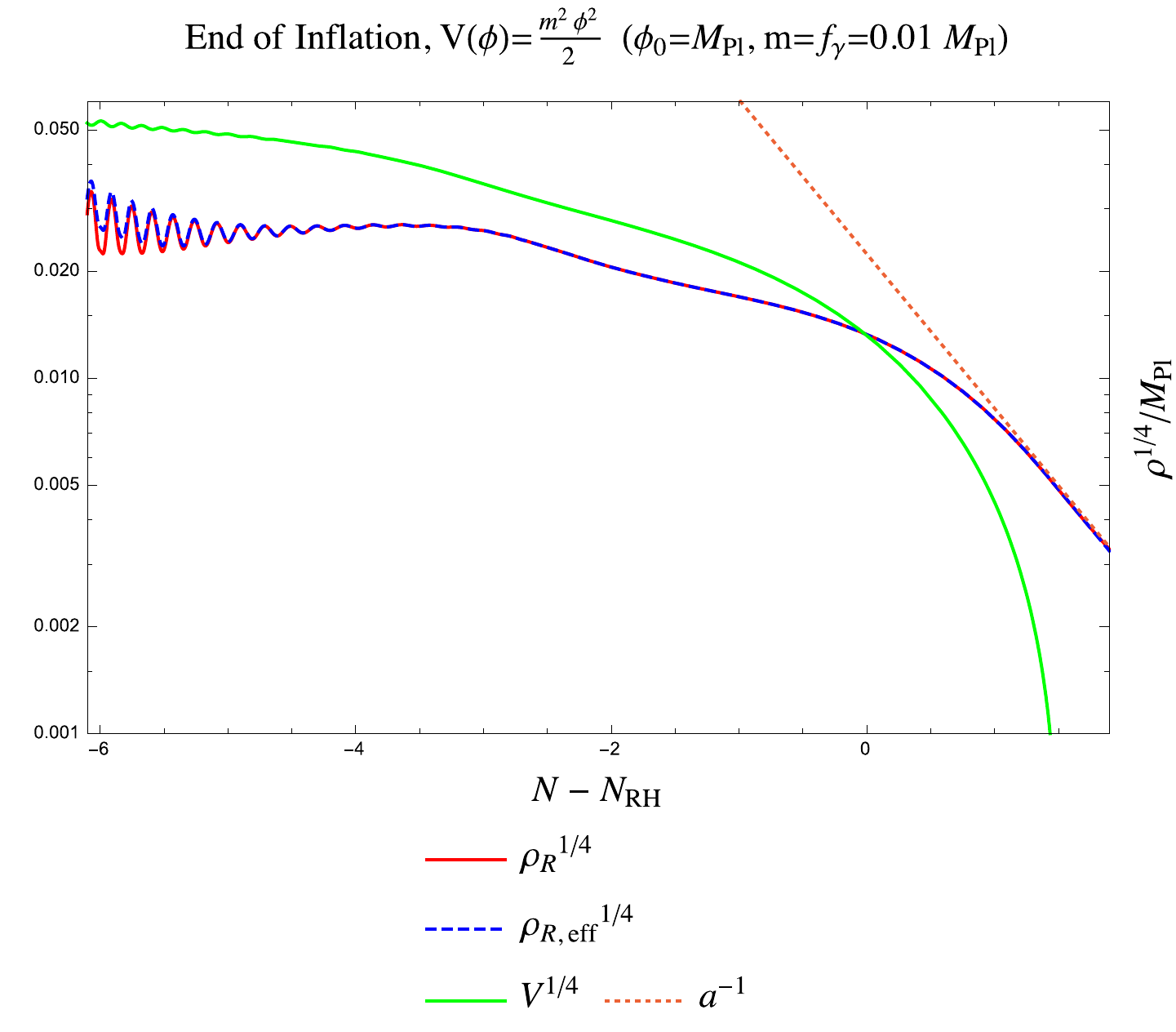}
    \end{tabular}
         \caption{In the left panel we plot the field evolution close to the end of inflation: the field slows down and stops in zero, as opposed to the standard reheating picture with oscillations around the minimum. In the right panel we show the energy density stored in the potential $V$ and in the gauge field. The red line corresponds to $\rho_R$ found using  the exact (renormalized) value eq.~(\ref{FLRW}),  while the blue dashed line corresponds to solving for eq.~(\ref{rhoReqinfl}). Note that after that the potential becomes subdominant, at $N=N_{RH}$, radiation domination starts and $\rho_R$ scales simply as $a^{-4}$.}
    \label{PlotRH}


\end{figure}

It is also important to stress how the end of inflation and reheating happens here. Actually there is no clear distinction between inflation and reheating, since radiation is produced continuously and its energy density stays roughly constant in time. The crucial difference is only that during inflation the potential energy dominates. We can find the field value $\phi_{RH}$ and also the temperature $T_{RH}$ at reheating just equating by definition the potential to the radiation energy density $\rho_{RH}$:
\begin{eqnarray}
V(\phi_{RH})= \rho_{RH} \, ,\qquad T_{RH}= \left( \frac{30 \rho_{RH}}{\pi^2 g_*}\right)^{1/4}
\end{eqnarray}
where, to define a temperature, we assumed that the radiation can actually reach quickly a thermal equilibrium (and $g_*$ is the number of degrees of freedom in the plasma which is formed).
Note that, remarkably, reheating here is fully fixed in this model and  $T_{RH}$ is already determined by the amount of radiation present during inflation (which is almost constant), in stark contrast with the usual slow-roll models with flat potentials, which have model-dependent reheating scenarios. We show one case, in fig.~\ref{PlotRH}, in which it is visible that the field is overdamped and just slows down approaching  $\phi_{RH}\approx 0$, which is consistent also with~\cite{Cheng:2015oqa}, while $\rho_R$ starts decreasing as $a^{-4}$ after $\phi=\phi_{RH}$. In some cases, if the coupling $1/f_\gamma$ is not very large, the field may also perform a few damped oscillations before relaxing at zero.
An estimate of $\rho_{RH}$, and so of the reheating temperature, is  given by $\rho_{RH}\approx {\cal B} \approx V' f_{\gamma}$, as can be seen from fig.~\ref{VphiplotFLRW}.

\begin{figure}
     \begin{tabular}{p{0.5\textwidth} p{0.5\textwidth}}
    \vspace{0pt} \    \includegraphics[width=0.53 \textwidth]{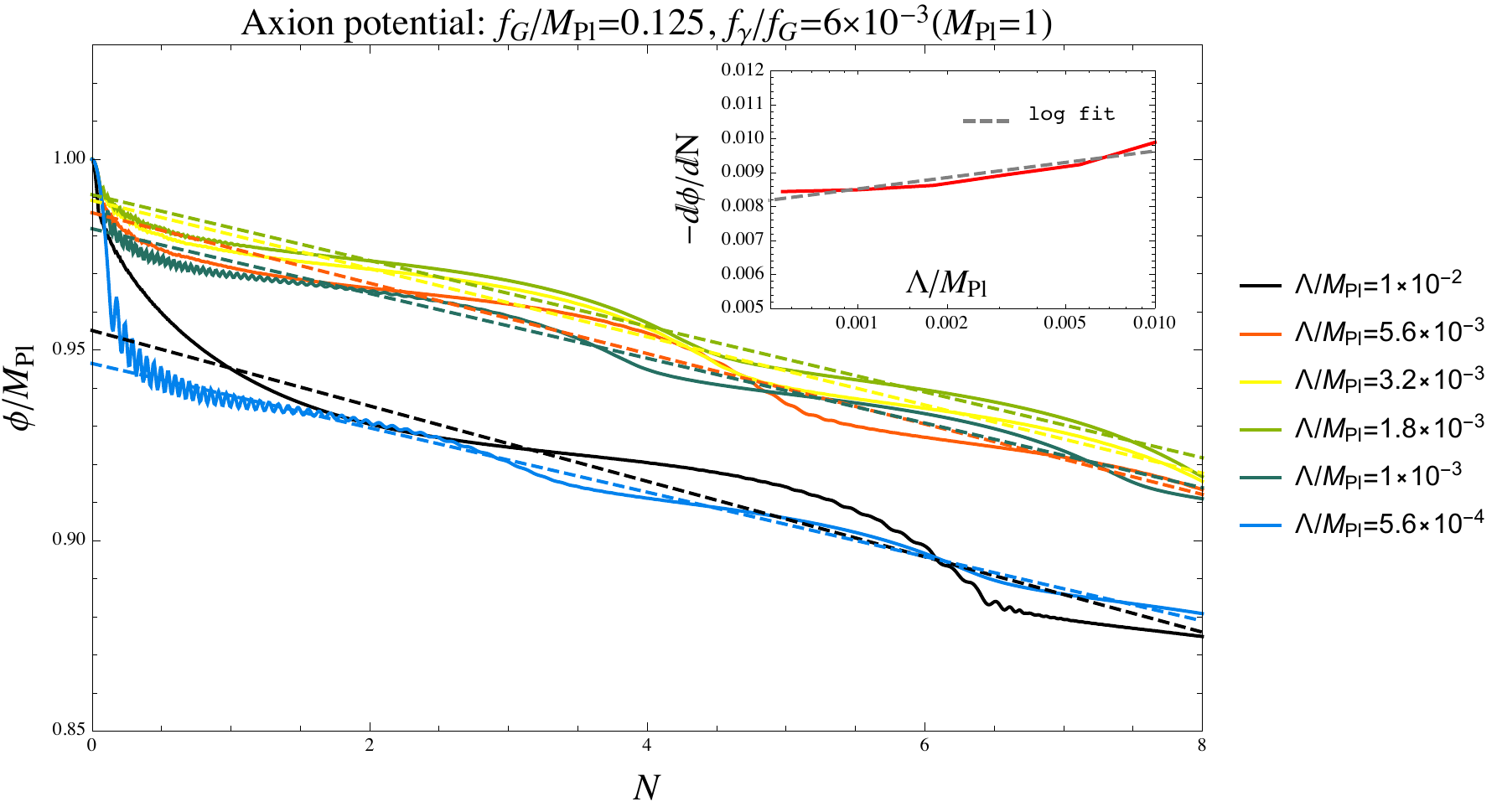} &
    \vspace{-2pt} \          \includegraphics[width=0.51 \textwidth]{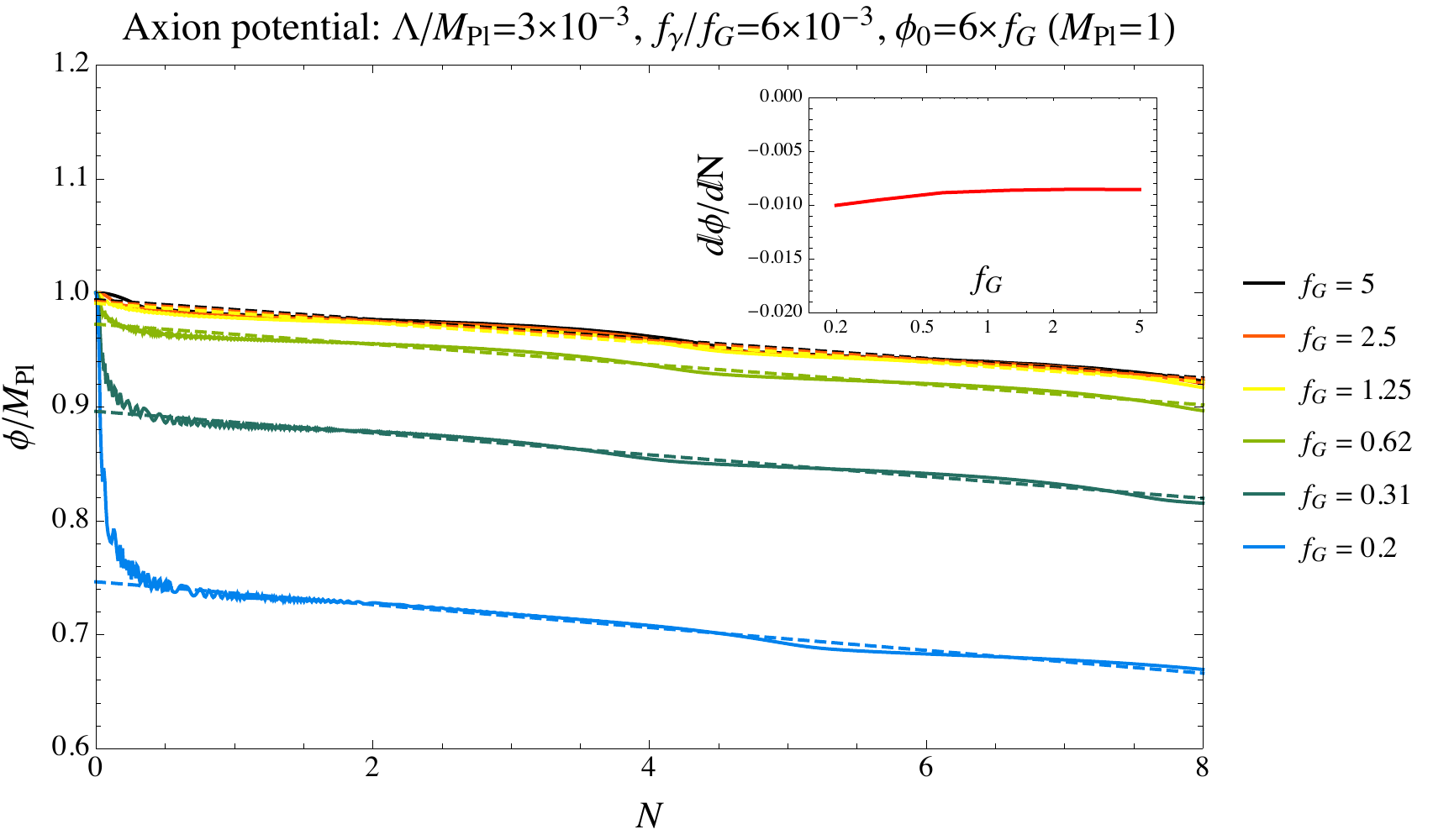}
                 \end{tabular}
\caption{We show here the $\phi$ behavior on an Axion potential varying the overall scale $\Lambda$ (left panel) and the scale $f_G$ (right panel) in eq.~\ref{Axion}. In both cases there is very weak dependence in the asymptotic late time behavior (in the small panels), while there is sensitivity only in the first efold of expansion. }
 \label{axionplots}
\end{figure}

It is easy to check that inflation with an axion-like potential, as in eq.~(\ref{Axion}), works practically in the same way as the quadratic case, as long as one starts in a generic point of the potential (we do not consider here fine-tuned situations with the initial $\phi$ very close to the maximum). We show some solutions for the axion potential in fig.~\ref{axionplots}. A particularly interesting case is the one of the QCD axion, with $\Lambda=\Lambda_{QCD}$ and so inflation happens at very low $H$. The reheating temperature would be at most of order $T_{RH} \lesssim \Lambda_{QCD}\approx {\cal O}(100 MeV)$. More precisely according to the above estimate $T_{RH}\approx \rho_{RH}^{1/4} \approx ( V' f_{\gamma} )^{1/4}\approx \Lambda_{QCD} \left(\frac{f_\gamma}{f_G} \right)^{1/4}$. This  can satisfy bounds from primordial nucleosynthesis only if $T_{RH}\gtrsim 5$ MeV. Unfortunately at such low values of $H$ our numerical solutions becomes exceedingly slow, and so we could not directly check if the scenario can work. Nonetheless as  already mentioned above, a major concern is to  check that during the transient phase, in the first efold of expansion, the field does not already fall down to zero. Using our flat space estimates this should be possible for very small $f_\gamma$, since $\dot{\phi}=\alpha f^2_\gamma$ and so we should impose that the field excursion $\Delta\phi\lesssim f_G$, where $\Delta\phi=\dot{\phi} \Delta t\approx \frac{\dot{\phi}}{H}\approx \frac{f^2_\gamma M_{Pl}} {\Lambda^2_{QCD}} $, which means $ f_\gamma\lesssim \left (\frac{f_G}{ M_{Pl}}\right)^{1/2} \Lambda_{QCD} $. Such a condition can be satisfied only if $f_G$ is several orders of magnitude larger than $M_{Pl}$, otherwise $f_\gamma$ turns out to be smaller than $\Lambda_{QCD}$ and therefore well below the experimental bounds~\cite{}.

Finally let us comment on the issue of density perturbations, which in such a scenario is non-trivial. First of all, one may wonder whether  the gauge modes generated by the background evolution may have a spectrum which can be relevant on large scales. The answer to this question is negative, as could already be seen from fig.~\ref{PBFLRW}. In fact a given $k$ mode is rapidly produced, when $k\approx \dot{\phi}/f_\gamma$, but afterwards its momentum redshifts and the growth  is not fast enough to overcome the $a^{-4}$ dilution. For this reason the spectrum of energy density is suppressed on large scales, roughly as $k^4$, and therefore its contribution to large scale density perturbations is negligible. For the same reason also magnetic fields on large scales would have negligible amplitude today\footnote{Note however that~\cite{Fujita:2015iga} invoked an {\it inverse cascade} scenario in which a helical magnetic field can transfer power from small to large scales, which should apply also in our case.}.

%


%
Nonetheless there is another source of density perturbation, which is due to the usual vacuum quantum fluctuations of the field $\phi$, coupled to a scalar part of the metric.  As far as we know the study of such perturbations in the backreaction-dominated regime has been performed so far only in a heuristic way~\cite{Anber:2009ua, Barnaby:2011vw, Barnaby:2012tk,Anber:2012du,Peloso:2016gqs} by looking at perturbations in $\phi$ and ignoring the metric components, while an equation for a gauge-invariant quantity, such as the comoving curvature perturbation has not been derived yet. Such a task is anyway highly non-trivial because of the presence of two components (the scalar and the photons), which exchange energy and because of the fact that the background itself has a contribution from the expectation value of the photon field, which is already an average over a collection of $k\neq 0$ modes, even in absence of fluctuations in $\phi$ and in the metric. We only mention here some qualitative features that we expect.

One can think of estimating the amplitude of the primordial curvature perturbation $A_\zeta$ in the so-called spatially flat gauge, where $\zeta = H \frac{\delta \rho}{\dot{\rho}} $ and where $\rho$ is  the total energy density and $\delta\rho$ its fluctuation; in usual single-field inflation this would give the well-known result
\begin{eqnarray}
A_\zeta \simeq   H \frac{V_\phi \delta \phi}{3 H \dot{\phi}^2}\approx \frac{H V_\phi }{\dot{\phi}^2}\approx \frac{H^2}{\dot{\phi}} \, , \label{usualzeta}
\end{eqnarray}
 where we have used the fact that $\phi$ has an almost flat spectrum of fluctuations with amplitude $\delta\phi\approx {\cal O}(H)$. In our case instead such an expression would become
\begin{eqnarray}
A_\zeta \simeq H \frac{\delta \rho}{\dot{\rho}}=  H \frac{V_\phi \delta \phi}{4 H \rho_R} \approx \frac{ V_\phi \delta\phi}{\rho_R} \label{Ampl}
\end{eqnarray}
where we have assumed that most of the energy density is stored in the potential of the $\phi$ field and so $\delta\rho\approx V_\phi \delta\phi$. The major challenge here is to know whether $\phi$ has a nearly flat spectrum and whether its amplitude is given by $H$ or some other quantity: one would have to write down and solve the coupled system of gauge field, metric and $\phi$ fluctuations around our background and quantize it to find the vacuum fluctuations.

Other subtle points are the following. In~\cite{Anber:2009ua} a term proportional to $\delta\dot{\phi}$ has been introduced by hand, on the basis that a $\delta{\dot\phi}$ should also dissipate its kinetic energy by exciting photons. However we argue that the situation is more involved and a possible dissipative effect has to depend on the physical momentum $p_{\rm phys}=p/a$ of the perturbation $\delta\dot{\phi}$, for two reasons. First the equation of motion for the vector potential ${\bf A}$ contains an extra term of the form $\nabla(\delta \phi) \times {\bf A}'$ which becomes important for large $p/a$, so that the analogy with the photons excited by a background $\phi(t)$ breaks down and we cannot expect a coherent excitation of photons  as a response to a $\delta\dot{\phi}$.
Second, even ignoring this, one would write down an equation of motion for $A_\pm$ similar to eq.~(\ref{coupled}), assuming a time dependent $\delta\dot{\phi}$ as a source. For instance, a massless field at $p/a\gg H$ would just have vacuum fluctuations with $\delta\phi_p=\frac{1}{a}\frac{e^{i p \tau}}{\sqrt{2 p}}$ behavior in Fourier  space. A given region of physical volume $L^3$ would have field oscillations in real space roughly of amplitude $1/L$ due to all modes with $p_{\rm phys}\lesssim 1/L$. Keeping the modes around $p_L/a\approx 1/L$ this translates in $\delta\phi \approx p_L/a \sin( p_L \tau)$, whose time derivative is (neglecting the expansion, in the regime $p_L/a\gg H$) roughly $\delta{\phi'}=p^2_L/a \cos( p_L \tau)$. This leads to an equation of motion of the form
\begin{eqnarray}
 A''_{\pm} + \left( k^2\mp  \frac{k  p^2_L }{a f_\gamma}  \cos(p_L \tau) \right) A_{\pm}=0\,  .
\end{eqnarray}
which is very similar to eq.~(\ref{flatquadratic}). It is easy to check that  only if $p_L/a \gg f_\gamma$ this equation can have large exponents for the Mathieu functions for some values of $k$. As a conclusion, when $p_L/a \lesssim f_\gamma$ we do not expect any backreaction  to be possible. 
Finally another concern is that in~\cite{Anber:2009ua} an ``inverse decay'' of the gauge modes created by the background was considered as  a source for $\delta\phi$. However, as long as the  $\delta\phi$ modes are not frozen completely there should be also a direct decay term $\delta\phi$ into gauge modes, and so the dynamics is likely to be much more complicated.

\begin{figure}
          \includegraphics[width=0.5 \textwidth]{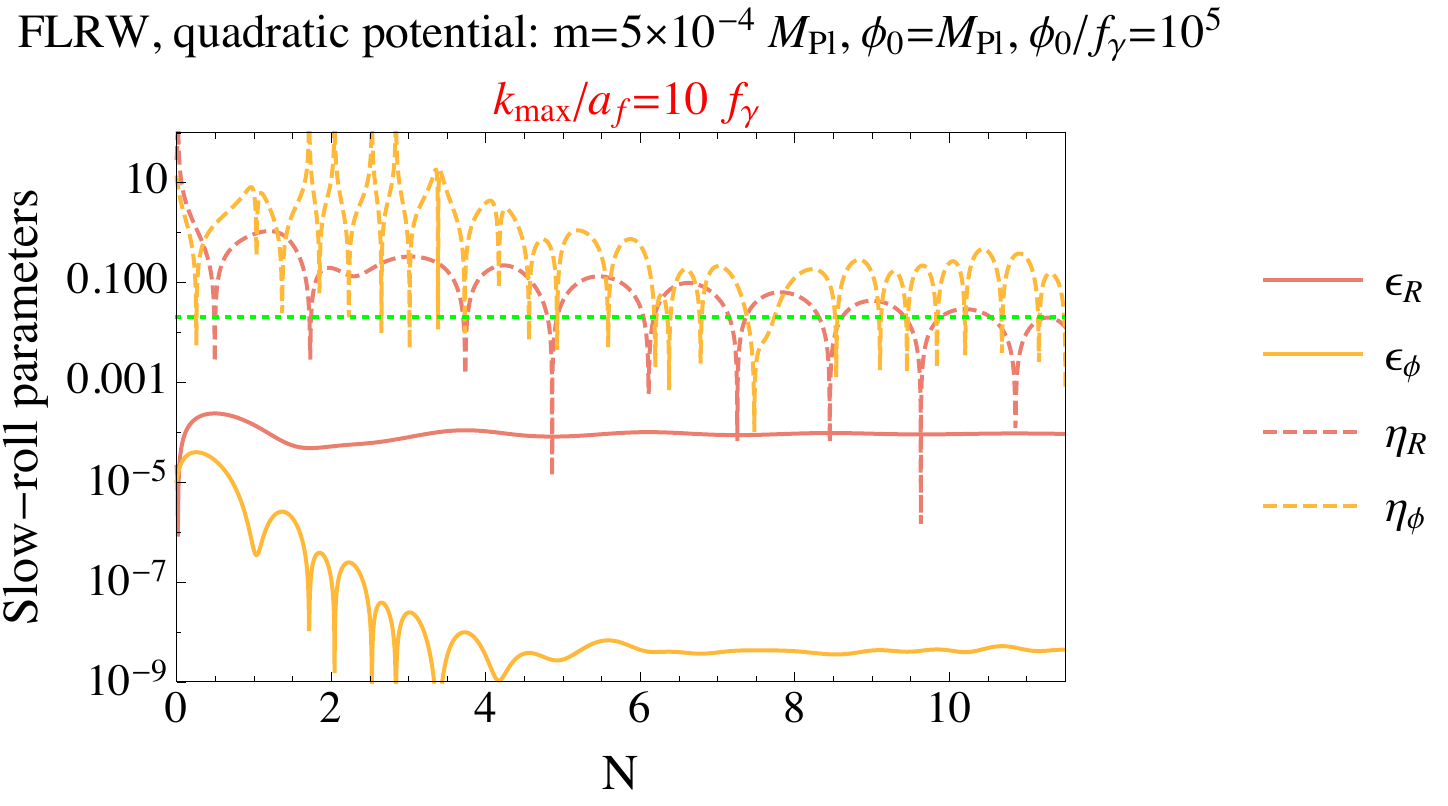}
           \includegraphics[width=0.5 \textwidth]{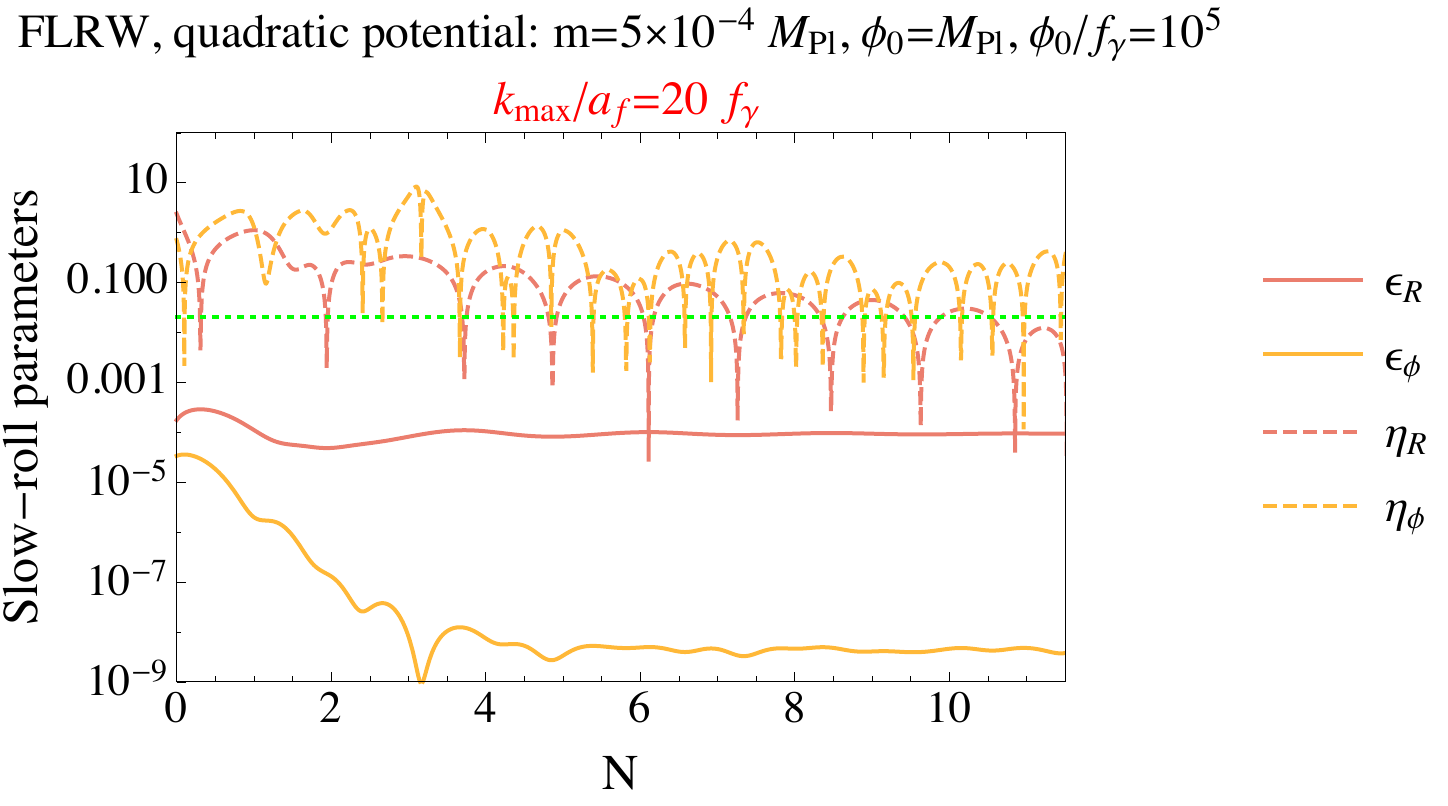}
     \caption{We plot the slow roll parameters defined in eqs.~\ref{slowroll}, for a quadratic potential $V=\frac{1}{2}m^2 \phi^2$: $\epsilon_R$ is always much larger than $\epsilon_\phi$, and they are both always small. In this case we could obtain smaller  $\eta_R$ and $\eta_\phi$ than in fig.~\ref{slowroll} using two ingredients: a very small $f_\gamma$ and also by including modes beyond the cutoff $f_\gamma$. The inclusion of such modes, while not justified in our effective treatment, turns out to provide a smoother evolution. In particular we have included modes up to $k_{max}/a_f=10 f_\gamma$ and $k_{max}/a_f=20 f_\gamma$ in the left and right panels, and  also we have implemented in both a step function which freezes the modes only when $k/a> 80 f_\gamma$. We show for visual reference the value 0.02 in green dashed.}
    \label{slowrolletasmall}
\end{figure}

We have therefore given arguments to conclude that a treatment of perturbations is very involved in this scenario and so we postpone this to future work.  
We only stress here that as a consequence of eq.~(\ref{Ampl}) the relevant slow-roll parameters are likely to be given by the ratio of $\rho_R$ over the Hubble rate, and its time variation. In analogy with the usual slow-roll parameters $\epsilon_\phi$ and $\eta_\phi$ defined from the derivatives of $\phi$, we can then define two slow-roll parameters for radiation:
\begin{eqnarray}
\epsilon_\phi &\equiv & \frac{\dot{\phi}^2}{2 M^2_{Pl} H^2}  \, , \qquad  \eta_\phi \equiv 2\epsilon_\phi+\frac{1}{2}\frac{d \log\epsilon_\phi}{dN} \, , \nonumber\\
\epsilon_R &\equiv  & \frac{2\rho_R}{3 M^2_{Pl} H^2} \,  , \qquad \eta_R \equiv 2\epsilon_R+\frac{1}{2}\frac{d \log\epsilon_R}{dN}
\label{slowroll}
\end{eqnarray}
One can define and compute also total slow-roll parameters as 
\begin{eqnarray}
\epsilon\equiv-\frac{\dot{H}}{H^2}=\epsilon_\phi+\epsilon_R\equiv \epsilon_R(1+\delta_\phi) \, , \,\,\,\,\,\,
\eta \equiv 2\epsilon+\frac{1}{2}\frac{d \log\epsilon}{dN}\simeq \eta_R+\frac{3}{2}\delta_\phi (\eta_R+\eta_\phi-\epsilon_\phi) \, ,
\end{eqnarray}
 where we have expanded at first order in the small parameter $\delta_\phi\equiv \frac{\dot{\phi}^2}{2 \rho_R}$, showing therefore that both total slow roll parameters are to a good approximation given by $\epsilon_R$ and $\eta_R$. Nonetheless it may be useful to plot the four slow-roll parameters for some illustrative cases. As already mentioned we leave for future work the calculation of the spectrum of perturbations, but it is reasonable that  $\epsilon_R$ and $\eta_R$ could be the relevant parameters. As it can be seen from fig.~\ref{slowroll} the $\epsilon_\phi$ and $\epsilon_R$ parameters are small, with $\epsilon_\phi\ll \epsilon_R$; however they have some superimposed oscillations that translate in values of $\eta_R$ and $\eta_\phi$ which are not very suppressed, ${\cal O}(0.1-1)$. Of course we know that decreasing $f_\gamma$ the  oscillations in $\phi$ have smaller amplitude, which would imply smaller $\eta$'s; however, as we have seen, when $f_\gamma$ is tiny the modes beyond the cutoff of the effective theory become not completely negligible. And in fact the only cases in which we could reach values of ${\cal O}(0.01)$  for $\eta_R$ were obtained also by including modes beyond the cutoff $f_\gamma$. The inclusion of such modes, while not justified in our effective treatment, turns out indeed to provide a smoother evolution. For instance in fig.~\ref{slowrolletasmall}  we have included modes up to $k_{max}/a_f=20 f_\gamma$ and $k_{max}/a_f=10 f_\gamma$,  and also we have implemented a step function which freezes the modes only when $k/a> 80 f_\gamma$. This show that in order to get a very smooth evolution it seems to be necessary to be able to treat also the modes beyond the cutoff, which means dealing with a more fundamental renormalizable UV complete model, which is however beyond the scope of this paper.

\begin{figure}
          \includegraphics[width=0.5 \textwidth]{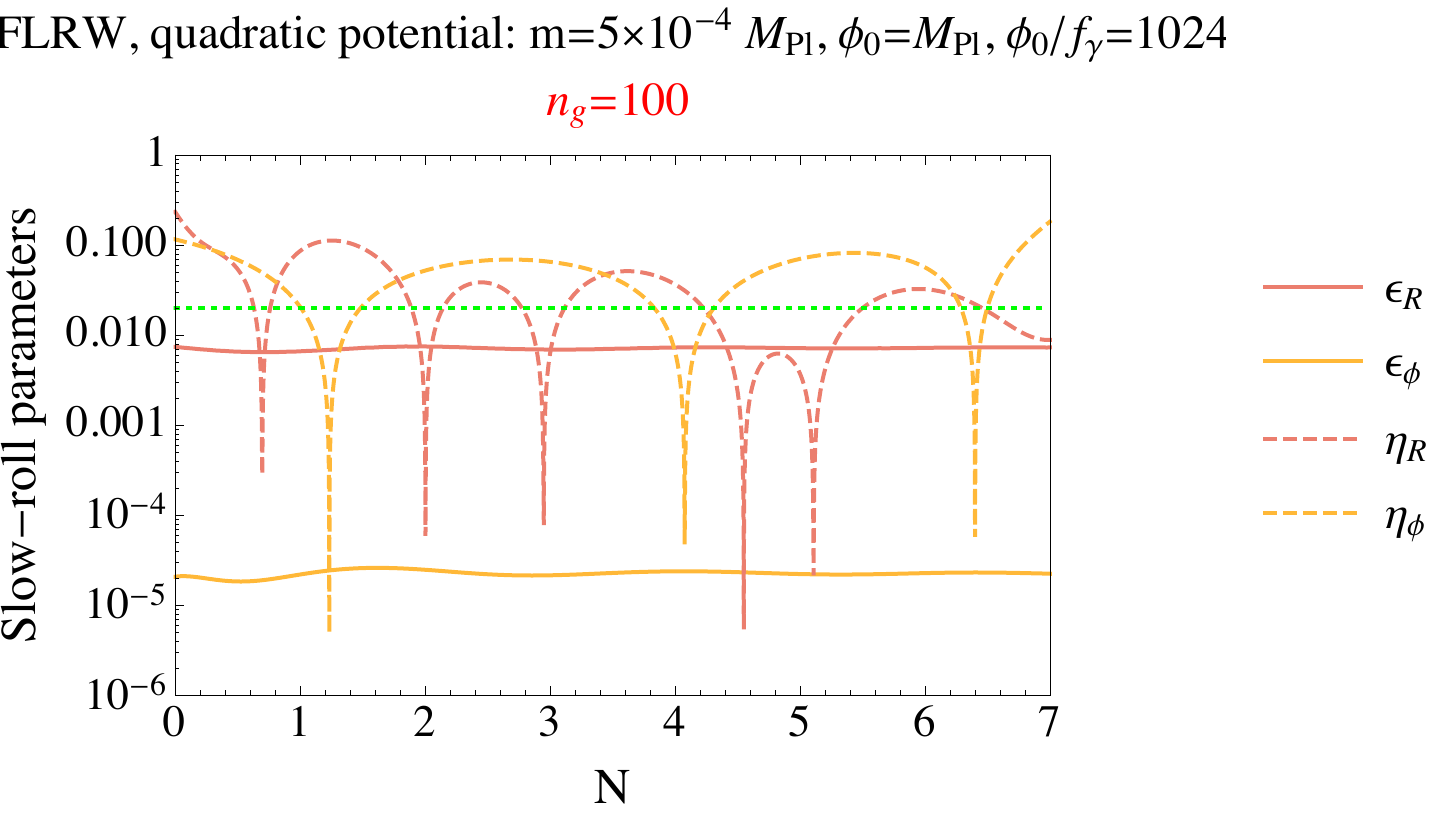}
           \includegraphics[width=0.5 \textwidth]{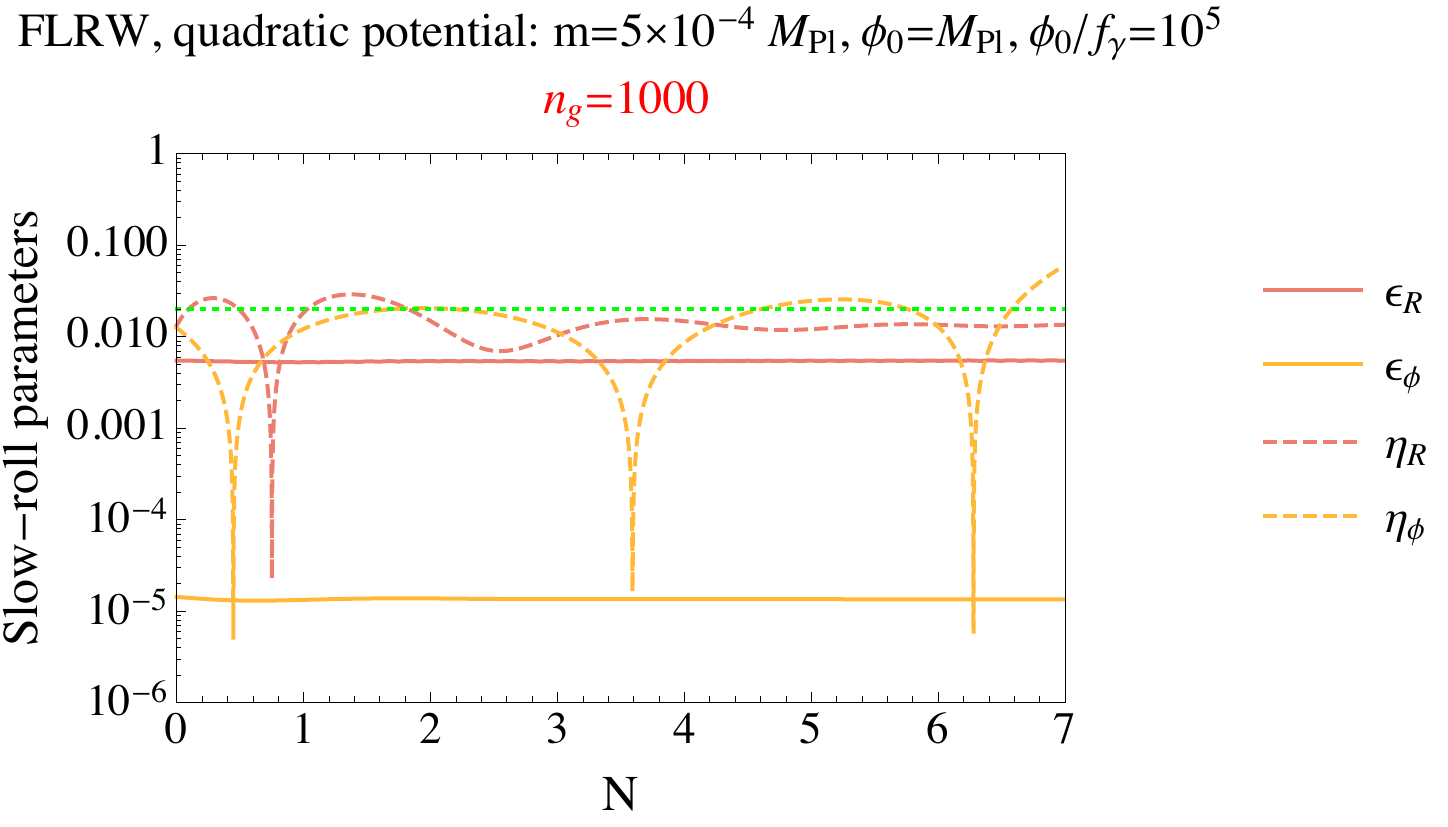}
     \caption{We plot the slow roll parameters defined in eqs.~\ref{slowroll}, for a quadratic potential $V=\frac{1}{2}m^2 \phi^2$: $\epsilon_R$ is always much larger than $\epsilon_\phi$, and they are both always small. In this case we could obtain smaller  $\eta_R$ and $\eta_\phi$ than in fig.~\ref{slowroll} using  a large number $n_g$ of identical species for the gauge fields. Here the cutoff is at $k_{max}/a_f=f_\gamma$, as usual. We show for visual reference the value 0.02 in green dashed.}
    \label{slowrolllargeng}
\end{figure}

Finally let us comment that another way to get a smoother evolution is to increase the multiplicity $n_g$ of gauge fields (note this was also invoked in~\cite{Anber:2009ua}, but for the different purpose of suppressing the amplitude of perturbations).
For instance in fig.~\ref{slowrolllargeng} if we consider a number of identical species $n_g\approx 10^2-10^3$ we get a smoother evolution and smaller and more stable values for $\eta_\phi$ and $\eta_R$.

 \begin{figure}
          \includegraphics[width=0.5 \textwidth]{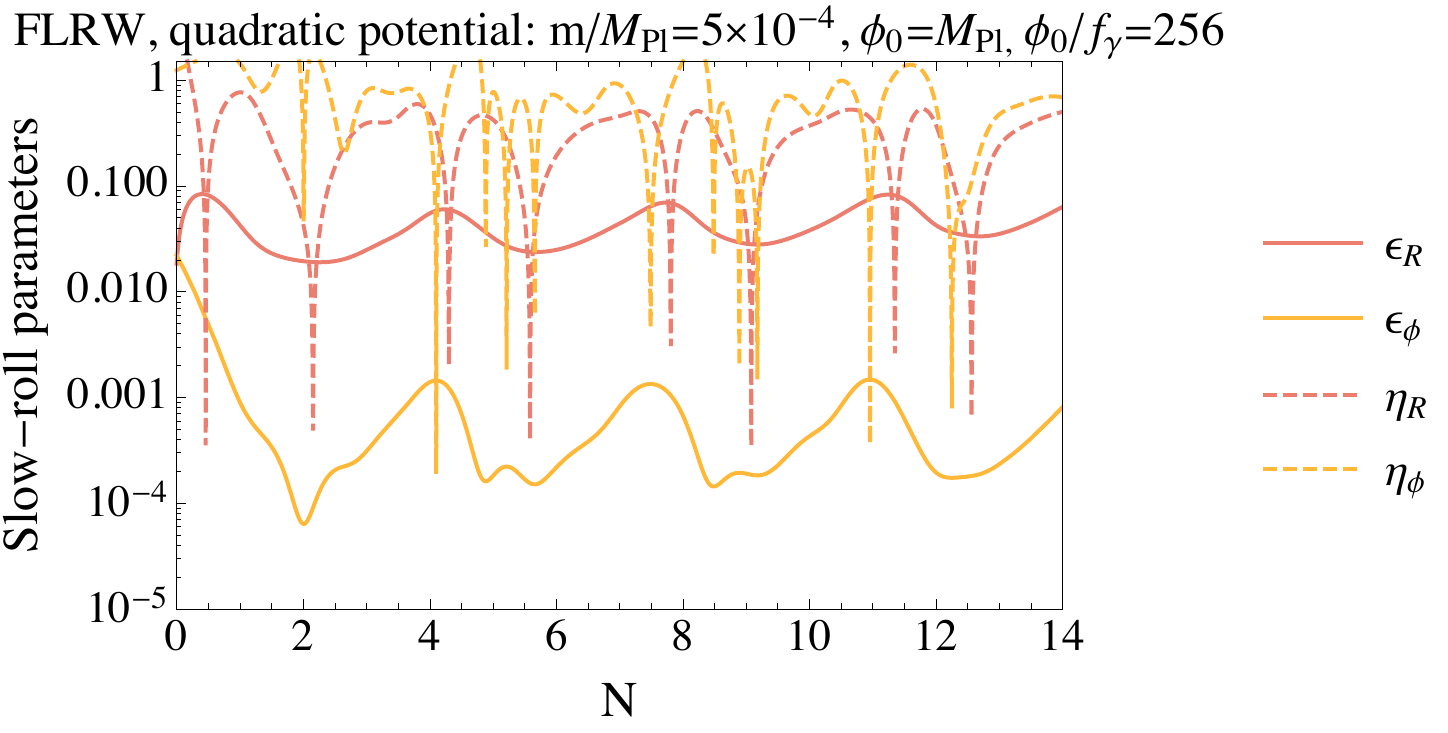}
           \includegraphics[width=0.5 \textwidth]{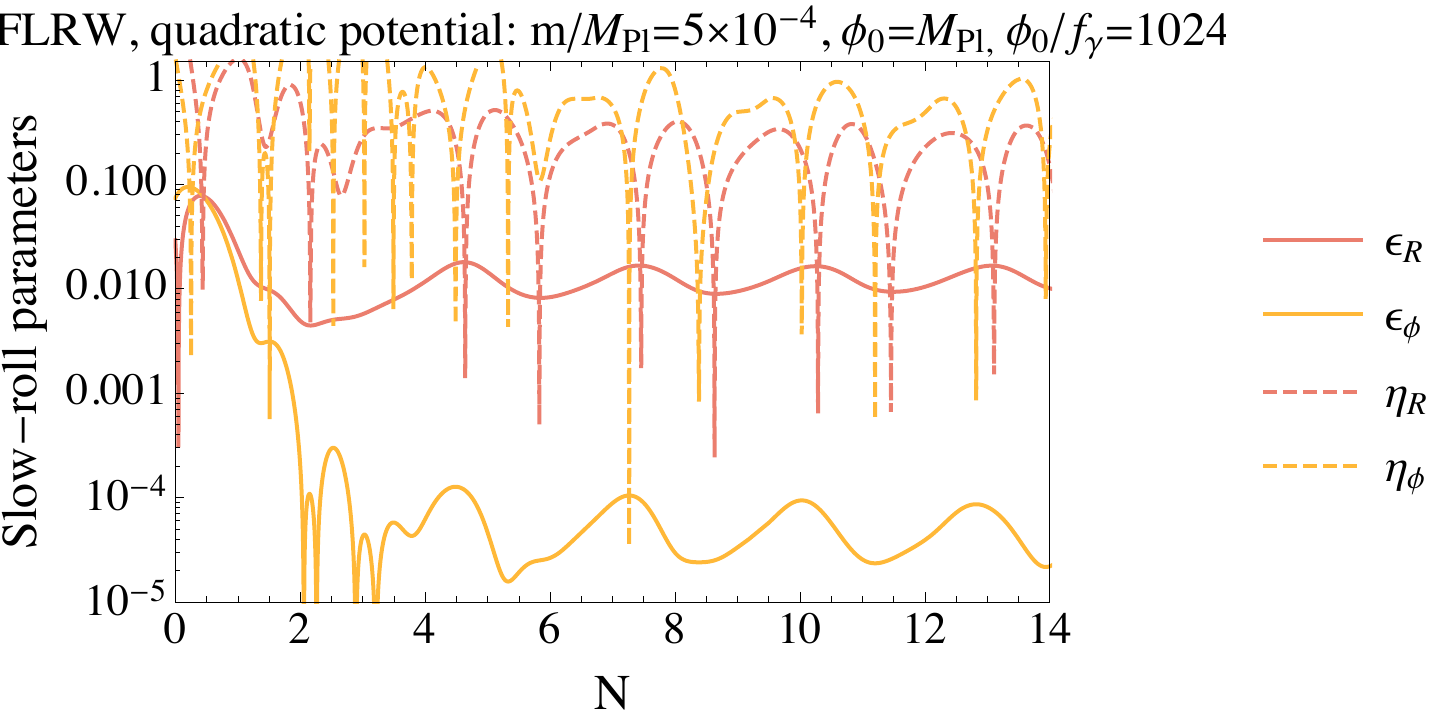}
     \caption{We plot the slow roll parameters defined in eqs.~\ref{slowroll}, for a quadratic potential $V=\frac{1}{2}m^2 \phi^2$: $\epsilon_R$ is always much larger than $\epsilon_\phi$, and they are both always small. Instead $\eta_R$ and $\eta_\phi$ are not extremely suppressed due to the oscillatory behavior of $\phi$ and in fact they are of ${\cal O}(0.1)$. Note also that all slow-roll parameters decrease for small $f_\gamma$. }
    \label{slowroll}
\end{figure}

Let us also stress that the background evolution implies that any additional light scalar field with mass $m\ll H$ would have an almost flat spectrum of fluctuations (but again with superimposed small oscillations) and therefore it  may be used to induce a flat spectrum of density perturbations, if it could dominate at any time the total energy density of the Universe, {\it i.e.} through the so-called curvaton mechanism~\cite{Lyth:2001nq}. 

Another important fact is that the tensor modes from inflation should have two contributions: a vacuum fluctuation with an amplitude $A_T=\frac{8}{M^2_{Pl}}\left(\frac{H}{2\pi}\right)^2$, and another one sourced by the gauge field~\cite{Cook:2011hg,Barnaby:2011qe,Anber:2012du,Sorbo:2011rz,Domcke:2016bkh}. Although we do not address such a calculation here, the generic prediction is again the presence of superimposed oscillations.  
In addition~\cite{Anber:2012du} there should be also parity violation, since the tensor modes produced by a source term should be chiral.

\section{Discussion} \label{discussion}
We have found that a scalar with a generic potential can  successfully achieve inflation at the background level through dissipation into gauge fields, as long as the coupling $1/f_\gamma$ is much larger than the inverse field excursion $1/\phi_0$, as in the case of an axion coupled to photons stronger than to gluons. This raises several questions and important issues.

First it is interesting to think whether such couplings might arise from a more complete theory. For instance a way to generate such a coupling is to integrate out a  fermion $\psi$, with $U(1)$ charge $g$ and with a $y \phi \bar{\psi} \gamma_5 \psi $ term, where $y$ is a coupling constant. This would mean that either $\phi$ is a pseudoscalar or alternatively, if $\phi$ is a scalar, that we have large CP violation. In any case this would generate an effective operator suppressed by the fermion mass  $m_f$, of the form $g^2 y \phi F \tilde{F}/m_f$, so that the new scale $f_\gamma$ would be given by $f_\gamma\equiv m_f/(g^2 y)$. However the very same coupling also  induces a contribution to $m_f$ of order $y \phi$. So in the end, barring cancellations with a tree level mass, we would typically have $f_\gamma=\phi/g^2$ and this could be much smaller than $\phi$ only if $g$ is large, which would represent a strong coupling regime. In order to avoid such a strongly coupled regime one could invoke also the presence of several fermionic species. Let us comment also that in such UV completions of our model it should be possible also to treat cases in which the tachyonic physical momenta are larger than $f_\gamma$. However this is beyond the scope of the present paper.

It is also interesting to think about other implications. For instance a curious phenomenon may arise already during inflation: if the $U(1)$ field is the actual electromagnetic field  it would actually produce pairs of electrons and positrons and other particles if the energy scale is high enough, so this might induce an even more complex dissipative dynamics. We do not expect however this to change the overall qualitative picture. In fact if such pairs are created during inflation they would be rapidly diluted away.

Finally we comment again on the production of large scale magnetic fields: while we have seen that $P_{\cal B}$ due to the background evolution is suppressed on large scales, it is still possible that large scale perturbations on $\phi$ may source large scale magnetic fields or that some later mechanism can transfer power to larger scales (see~\cite{Fujita:2015iga}). 
Another interesting feature is the fact that we are producing a nonzero $\vec{E}\cdot \vec{B}$ and this might in principle be  a source of parity violation in the early Universe~\cite{Sorbo:2011rz}, although again such an effect seems likely to be also suppressed at large scales. We postpone however such analysis to future work.

\section{Conclusions} \label{conclusions}

In this paper we have solved explicitly the dynamics for the background in a system in which inflation can take place without the need for a flat potential. This happens through an effective coupling of the form $\phi F \tilde{F}/f_\gamma$, as  proposed already in~\cite{Anber:2009ua}, which produces an instability in the gauge fields that can be viewed as a dissipation that slows down the $\phi$ field. We have studied the onset of such instability starting from a static configuration of the scalar field by numerically solving the coupled system of scalar and gauge field. We have shown that, even in flat spacetime, the mechanism is efficient as long as the instability is faster than the free field evolution, which can happen if the scale $f_\gamma$ is smaller than the field excursion $\phi_0$. Then we have extended this to the FLRW case, showing that if the mechanism can freeze the field value for at least about one efold, then the system enters the regime studied in~\cite{Anber:2009ua} with  the energy density staying nearly constant and the universe inflating. In particular $d\phi/dN\approx  c f_\gamma$, where $c$ is a numerical factor, which depends only logarithmically on the potential and on the Hubble scale and so, as long as the mass scale $f_\gamma$ is much smaller than the field excursion $\phi_0$, the mechanism is highly efficient, providing easily more than the required 60 efolds of inflation. The mechanism is appealing, since it does not require any fine-tuned potential, nor superplanckian field excursions, and the basic requirements are: an equation of motion which violates $CP$ (and therefore $T$), since it can contain a term proportional to $\dot{\phi}$, and the dissipation of energy into massless degrees of freedom, so that the friction is efficient for very long time.
We have studied the system cutting off modes with momentum $k>f_\gamma$: this is a good approximation except at extremely small $f_\gamma$ ({\it i.e.} very large friction), in which case a full knowledge of the theory seems necessary beyond the effective theory. An important feature of the numerical solutions is the presence of oscillations in the background solution, with a typical period of about 4-5 efolds and an amplitude which is linear in $f_\gamma$ and which is less than ${\cal O}(0.1)\%$ if a total number of efolds $N\gtrsim 60$ is required. 

We have also argued that an axion field can have such properties if the coupling to photons $1/f_\gamma$ is much larger than the one to gluons $1/f_G$, since the latter scale determines the maximal field excursion, $f_G\sim \phi_0$.

We have left the calculation of the spectrum and the non-gaussianity of the perturbations for future work.  This is a crucial point since according to~\cite{Anber:2009ua} this model  has a too large amplitude of the  power spectrum $A_\zeta$, unless a large number of species for the gauge fields is considered, and another concern is the size of the nonlinearity parameters $f_{NL}$, which was shown to grow  large with $\xi$, at least in absence of backreaction~\cite{Barnaby:2011vw,Barnaby:2011qe}. However we think that the dynamics of the perturbations in the backreaction dominated regime should be treated more carefully: the action of a gauge-invariant variable (such as the comoving curvature perturbation $\zeta$) should take into account of the non-trivial background in which an energy density $\rho_R$ in radiation is present, so that $\dot{\rho} \approx 4 H \rho_R$. So, an action to second and third order for $\zeta$ in the presence of $\phi$, $A_\mu$ and other metric perturbations should be consistently written. Also, we have argued that the effect of gauge fields on the perturbations $\delta\phi$ cannot be treated on all scales as a dissipative term proportional to $\delta\dot{\phi}$, as in~\cite{Anber:2009ua}, due to the presence of gradients of $\delta\phi$ and due to the fact that such a dissipation should be efficient only on scales much larger than $1/f_\gamma$. Moreover we have pointed out that the interaction term between $\delta\phi$ and the gauge field should contain not only ``inverse decays'', but also direct decays. Even if we did not solve the challenging task of solving for the perturbations, we have argued that new slow-roll parameters $\epsilon_R$ and $\eta_R$, proportional to $\rho_R/H^2$ and its time variations may be relevant for the evolution of scalar fluctuations. For this purpose we have shown that while $\epsilon_R$ can be rather small (though much bigger than the usual $\epsilon_\phi$ parameter), it is more difficult to make $\eta_R$ less than ${\cal O}(0.1)$. The latter may be achieved at tiny values of $f_\gamma$, but at the price of including modes of about a factor of 10 beyond the cutoff $k/a\lesssim {\cal O}(10) f_\gamma$. This is out of the regime of validity of our effective coupling, but nonetheless it may be an indication that in a UV complete theory a smooth evolution can be achieved. Alternatively we have shown that in the presence of a large number of gauge fields, ${\cal O}(10^2-10^3)$, $\eta_R$ can be strongly reduced.

We have also argued that the addition of a curvaton field might imprint a flat spectrum of perturbations, in case the spectrum of $\zeta$ turns out to be negligible on large scales, since such a field would have an almost flat spectrum of amplitude $H$ in this inflationary background.
Finally we stress again that the background always has small superimposed oscillations in the evolution of $H$ and thus, whatever is the mechanism of production of cosmological scalar and tensor perturbations, there should necessarily be an imprint on the observable density and tensor fluctuations at late times, which can constitute a distinctive generic feature of such a scenario. 

${}$\linebreak
\emph{\textbf{Acknowledgments.}}
We thank Guillermo Ballesteros, Filippo Vernizzi, Guido D'Amico,  Jaume Garriga, Marco Peloso, Yuko Urakawa, Jorge Nore\~na, Giovanni Villadoro, Federico Mescia, Domenec Espriu, Denis Comelli and Cristiano Germani for many useful discussions. AN is supported by the grants EC FPA2010-20807-C02-02, AGAUR 2009-SGR-168.
KT has been supported by a Marie Sk\l{}odowska-Curie Individual Fellowship of the European Commission's Horizon 2020 Programme under contract number 655279 ResolvedJetsHIC.

\bibliographystyle{unsrt}
\bibliography{gaugeinflation}

\begin{thebibliography}{10}

\bibitem{Starobinsky:1980te}
Alexei~A. Starobinsky.
\newblock {A New Type of Isotropic Cosmological Models Without Singularity}.
\newblock {\em Phys. Lett.}, B91:99--102, 1980.

\bibitem{Guth:1980zm}
Alan~H. Guth.
\newblock {The Inflationary Universe: A Possible Solution to the Horizon and
  Flatness Problems}.
\newblock {\em Phys. Rev.}, D23:347--356, 1981.

\bibitem{Linde:1981mu}
Andrei~D. Linde.
\newblock {A New Inflationary Universe Scenario: A Possible Solution of the
  Horizon, Flatness, Homogeneity, Isotropy and Primordial Monopole Problems}.
\newblock {\em Phys. Lett.}, B108:389--393, 1982.

\bibitem{Berera:1995ie}
Arjun Berera.
\newblock {Warm inflation}.
\newblock {\em Phys. Rev. Lett.}, 75:3218--3221, 1995.

\bibitem{Anber:2009ua}
Mohamed~M. Anber and Lorenzo Sorbo.
\newblock {Naturally inflating on steep potentials through electromagnetic
  dissipation}.
\newblock {\em Phys. Rev.}, D81:043534, 2010.

\bibitem{Cheng:2015oqa}
Shu-Lin Cheng, Wolung Lee, and Kin-Wang Ng.
\newblock {Numerical study of pseudoscalar inflation with an axion-gauge field
  coupling}.
\newblock {\em Phys. Rev.}, D93(6):063510, 2016.

\bibitem{Tkachev:1986tr}
I.~I. Tkachev.
\newblock {Coherent scalar field oscillations forming compact astrophysical
  objects}.
\newblock {\em Sov. Astron. Lett.}, 12:305--308, 1986.
\newblock [Pisma Astron. Zh.12,726(1986)].

\bibitem{DiVecchia:1980yfw}
P.~Di~Vecchia and G.~Veneziano.
\newblock {Chiral Dynamics in the Large n Limit}.
\newblock {\em Nucl. Phys.}, B171:253--272, 1980.

\bibitem{Ade:2015lrj}
P.~A.~R. Ade et~al.
\newblock {Planck 2015 results. XX. Constraints on inflation}.
\newblock 2015.

\bibitem{Meerburg:2013dla}
P.~Daniel Meerburg and David~N. Spergel.
\newblock {Searching for oscillations in the primordial power spectrum. II.
  Constraints from Planck data}.
\newblock {\em Phys. Rev.}, D89(6):063537, 2014.

\bibitem{Fujita:2015iga}
Tomohiro Fujita, Ryo Namba, Yuichiro Tada, Naoyuki Takeda, and Hiroyuki
  Tashiro.
\newblock {Consistent generation of magnetic fields in axion inflation models}.
\newblock {\em JCAP}, 1505(05):054, 2015.

\bibitem{Barnaby:2011vw}
Neil Barnaby, Ryo Namba, and Marco Peloso.
\newblock {Phenomenology of a Pseudo-Scalar Inflaton: Naturally Large
  Nongaussianity}.
\newblock {\em JCAP}, 1104:009, 2011.

\bibitem{Barnaby:2012tk}
Neil Barnaby, Ryo Namba, and Marco Peloso.
\newblock {Observable non-gaussianity from gauge field production in slow roll
  inflation, and a challenging connection with magnetogenesis}.
\newblock {\em Phys. Rev.}, D85:123523, 2012.

\bibitem{Anber:2012du}
Mohamed~M. Anber and Lorenzo Sorbo.
\newblock {Non-Gaussianities and chiral gravitational waves in natural steep
  inflation}.
\newblock {\em Phys. Rev.}, D85:123537, 2012.

\bibitem{Peloso:2016gqs}
Marco Peloso, Lorenzo Sorbo, and Caner Unal.
\newblock {Rolling axions during inflation: perturbativity and signatures}.
\newblock 2016.

\bibitem{Lyth:2001nq}
David~H. Lyth and David Wands.
\newblock {Generating the curvature perturbation without an inflaton}.
\newblock {\em Phys. Lett.}, B524:5--14, 2002.

\bibitem{Cook:2011hg}
Jessica~L. Cook and Lorenzo Sorbo.
\newblock {Particle production during inflation and gravitational waves
  detectable by ground-based interferometers}.
\newblock {\em Phys. Rev.}, D85:023534, 2012.
\newblock [Erratum: Phys. Rev.D86,069901(2012)].

\bibitem{Barnaby:2011qe}
Neil Barnaby, Enrico Pajer, and Marco Peloso.
\newblock {Gauge Field Production in Axion Inflation: Consequences for
  Monodromy, non-Gaussianity in the CMB, and Gravitational Waves at
  Interferometers}.
\newblock {\em Phys. Rev.}, D85:023525, 2012.

\bibitem{Sorbo:2011rz}
Lorenzo Sorbo.
\newblock {Parity violation in the Cosmic Microwave Background from a
  pseudoscalar inflaton}.
\newblock {\em JCAP}, 1106:003, 2011.

\bibitem{Domcke:2016bkh}
Valerie Domcke, Mauro Pieroni, and Pierre Bin\'etruy.
\newblock {Primordial gravitational waves for universality classes of
  pseudoscalar inflation}.
\newblock 2016.

\end{thebibliography}

\end{document}